\documentclass[lettersize,journal]{IEEEtran}

\usepackage{array}
\usepackage[caption=false,font=normalsize,labelfont=sf,textfont=sf]{subfig}
\usepackage{textcomp}
\usepackage{stfloats}
\usepackage{url}
\usepackage{verbatim}
\usepackage{cite}
\usepackage{tikz}
\usepackage[english]{babel}
\usepackage{blindtext}
\usepackage{xspace}
\usepackage{graphicx}
\usepackage{multirow}
\usepackage[inline]{enumitem}
\graphicspath{{figures/}}
\usepackage{_macros}
\usepackage{algpseudocode}
\usepackage{algorithm}
\usepackage{amsmath}
\usepackage{amsfonts}
\usepackage{xurl}
\usepackage{hyperref}
\usepackage{breakurl}

\begin{document}

\title{Enhancing IoT Communication and Localization via Smarter Antenna}

\newcommand{\name}{Wi-Pro\xspace}
\newcommand{\hf}[1]{\authorcomment{PURPLE}{H}{#1}} 
\newcommand{\oa}[1]{\authorcomment{RED}{O}{#1}}
\newcommand{\ptr}[1]{\authorcomment{blue}{P}{#1}}

\author{Tianxiang Li,
Haofan Lu,
and~Omid~Abari
\thanks{Tianxiang Li did this work at the University of California at Los Angeles (UCLA), before joining Meta Platforms.}
\thanks{Omid Abari and Haofan Lu are with the Department of Computer Science, University of California at Los Angeles (UCLA), CA 90095, USA (email: omid@cs.ucla.edu, haofan@cs.ucla.edu)}
}


\maketitle
\footnote{This work has been submitted to the IEEE IoT Journal for possible publication. Copyright may be transferred without notice, after which this version may no longer be accessible.}
\begin{abstract}
The convergence of sensing and communication functionalities is poised to become a pivotal feature of the sixth-generation (6G) wireless networks. This vision represents a paradigm shift in wireless network design, moving beyond mere communication to a holistic integration of sensing and communication capabilities, thereby further narrowing the gap between the physical and digital worlds. While Internet of Things (IoT) devices are integral to future wireless networks, their current capabilities in sensing and communication are constrained by their power and resource limitations. On one hand, their restricted power budget limits their transmission power, leading to reduced communication range and data rates. On the other hand, their limited hardware and processing abilities hinder the adoption of sophisticated sensing technologies, such as direction finding and localization. In this work, we introduce \name, a system which seamlessly integrates today's WiFi protocol with smart antenna design to enhance the communication and sensing capabilities of existing IoT devices. This plug-and-play system can be easily installed by replacing the IoT device's antenna. \name seamlessly integrates smart antenna hardware with current WiFi protocols, utilizing their inherent features to not only enhance communication but also to enable precise localization on low-cost IoT devices. Our evaluation results demonstrate that \name achieves up to 150\% data rate improvement, up to five times range improvement, accurate direction finding, and localization on single-chain IoT devices.
\end{abstract}


\section{Introduction}
\label{introduction}
\IEEEPARstart{T}{he} evolution of telecommunication networks into the sixth generation (6G) heralds a transformative era where the integration of sensing and communication functions becomes a foundational pillar. 
This paradigm shift aims to not only transmit data with better speed and efficiency but also to perceive the environment and context. The great variety of Internet of Things (IoT) devices serves as a pivotal infrastructure in the convergence of sensing and communication. Unfortunately, most IoT devices today have limited power and processing resources, which limits their communication and sensing capabilities. Specifically, on the one hand, due to the limited power budget, IoT devices typically have lower transmission power than devices with constant power supply, which limits their communication range and data rate. On the other hand, IoT devices cannot benefit from advanced communication and sensing techniques, such as MIMO and phased array, as they require complex software and hardware unavailable on low-cost IoT devices.  

\begin{figure}[t!]
    \centering
    \subfloat[Enhancing WiFi Communication with \textit{\name}]{
        \includegraphics[width=0.3\textwidth]{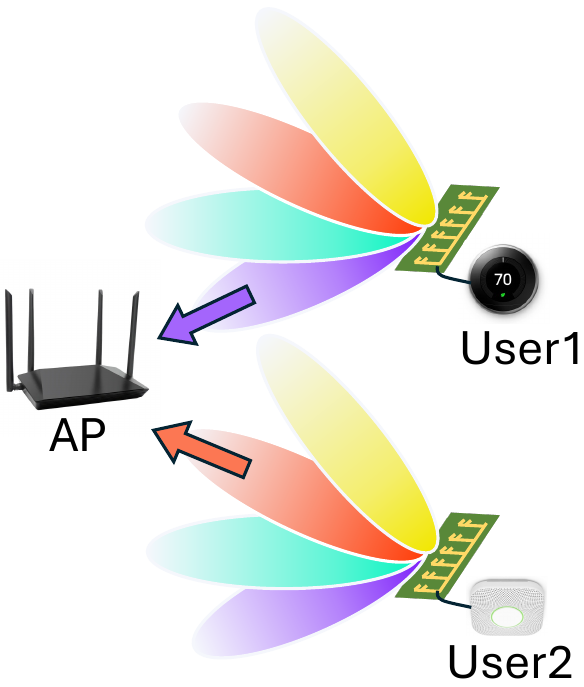}
        \captionsetup{width=1.6\textwidth}
        \label{fig:wipro_comm}
    }
    \hfill
    \subfloat[Enhancing WiFi Localization with \textit{\name}]{
        \includegraphics[width=0.3\textwidth]{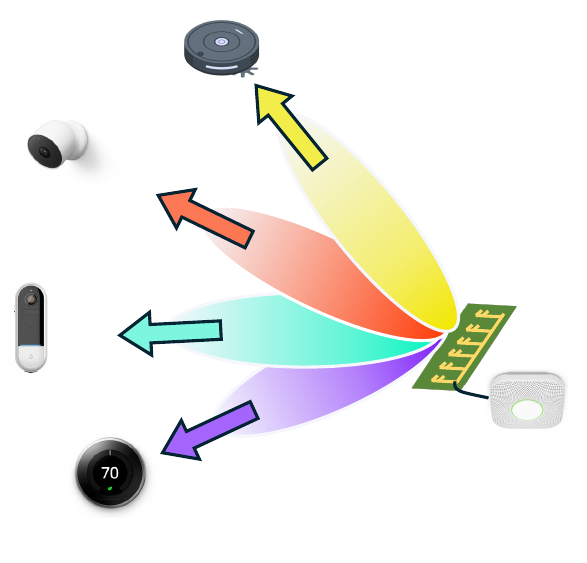}
        \captionsetup{width=1.6\textwidth}
        \label{fig:wipro_loc}
    }
    \caption{\name Overview. \textmd{(a) \name amplifies the signal transmitted or received by each IoT device in the direction of the access point and in the frequency used by that device to enhance the data rate and communication range. 
       (b) \name enables an IoT device with a single transceiver chain to localize other WiFi devices seamlessly.}}
    \label{fig:overview}
\end{figure}

     


In this paper, we propose \name, a simple while effective system that enhances the communication and sensing capabilities of IoT devices with a unified design. The core of \name is a new type of WiFi antenna that integrates seamlessly with today's WiFi protocols. 
Figure~\ref{fig:overview} shows the two use cases of \name. In the first scenario (Figure~\ref{fig:wipro_comm}), \name concentrates an IoT device's signal power toward the Access Point (AP) in order to enhance the Signal-to-Noise Ratio (SNR) and improve the range and data rate for communication. Different from conventional directional antennas, \name has a wide field of view, which allows it to always point its beam to the AP. Note that \name achieves this \textit{without} requiring complex hardware and processing algorithms such as MIMO and phased array; therefore, can be easily applied to resource-constrained IoT devices. In the second scenario (Figure~\ref{fig:wipro_loc}), \name enables an IoT device to localize other WiFi devices. 
This can enable a great variety of new applications. For instance, in smart homes, smart thermostats could adjust the temperature based on the location of occupants in the home; lighting systems could turn off lights in unoccupied rooms or adjust the brightness.
Conventional localization methods either require the trilateration/triangulation of multiple devices to locate a target device or require a single device with multiple transceiver chains to measure the angle and distance to the target device. Unfortunately, most of IoT devices have only a single transceiver chain; therefore, they cannot utilize these methods. \name solves this problem using a novel antenna structure that allows a single-chain IoT device to measure the direction of other WiFi devices and eventually localize the target device.

At a high level, \name leverages a passive beamforming antenna whose beam directions depend on the frequencies of the signal. We show that integrating this unique property with the latest WiFi protocols enhances the communication and sensing capability of IoT devices significantly.
In particular, we first showcase how \name enhances IoT communication by integrating with Orthogonal Frequency-Division Multiple Access (OFDMA) in today's WiFi protocol. 
Specifically, OFDMA allocates different chunks of the subcarriers to different user devices. \name leverages this frequency-domain resource allocation to naturally form a beam toward the AP. In other words, \name's antenna provides an additional amplification to the signal it receives and transmits within the frequency range allocated to the IoT device. In this way, \name improves the signal strength and enhances the communication of IoT devices without mofidying the existing WiFi protocol, or requiring any complex, power-hungry hardware. Then, we show that \name can enable localization for single-chain IoT devices by leveraging the inherent WiFi probing mechanism. WiFi devices send probe request packets on each frequency channel periodically. For a particular device sending probe request packets while scanning through all the frequency channels, \name amplifies the received signal power for some of the probe request packets if they are aligned with the direction of the beam. This allows an IoT device to measure the signal angle-of-arrival (AoA) and infer the direction of the target device by simply sniffing on each frequency channel. We further measure the distance to the target device using the round-trip time method to enable localization.


We implement a prototype of \name and conduct extensive experiments to evaluate its performance in terms of communication SNR and localization accuracy. The evaluation results exhibit that \name improves the throughput of an IoT device by 10$\sim$120\%, extends the communication range by more than three times, and enables direction finding with an average precision of 1.1 degrees.

In summary, we make the following contributions
\begin{itemize}

    \item We introduce \name, the first system that seamlessly integrates today's WiFi protocol with smart antenna design to enhance the communication and localization capability of resource-constrained IoT devices.

    \item \name enhances the communication range and data rate by passively strengthening the signal power based on the location of the IoT device.

    \item \name enables the device localization on low-cost IoT devices with a single transceiver chain.  

    \item We build a prototype of \name and demonstrate significant improvement in communication range (up to five times) and data-rate  (up to 150\%), as well as achieving accurate localization.
    
\end{itemize}

\section{Background}
\label{sec:background}
In this section, we provide some background information on topics related to \name.

\begin{figure}[t!]
    \centering
    \includegraphics[width=7cm]{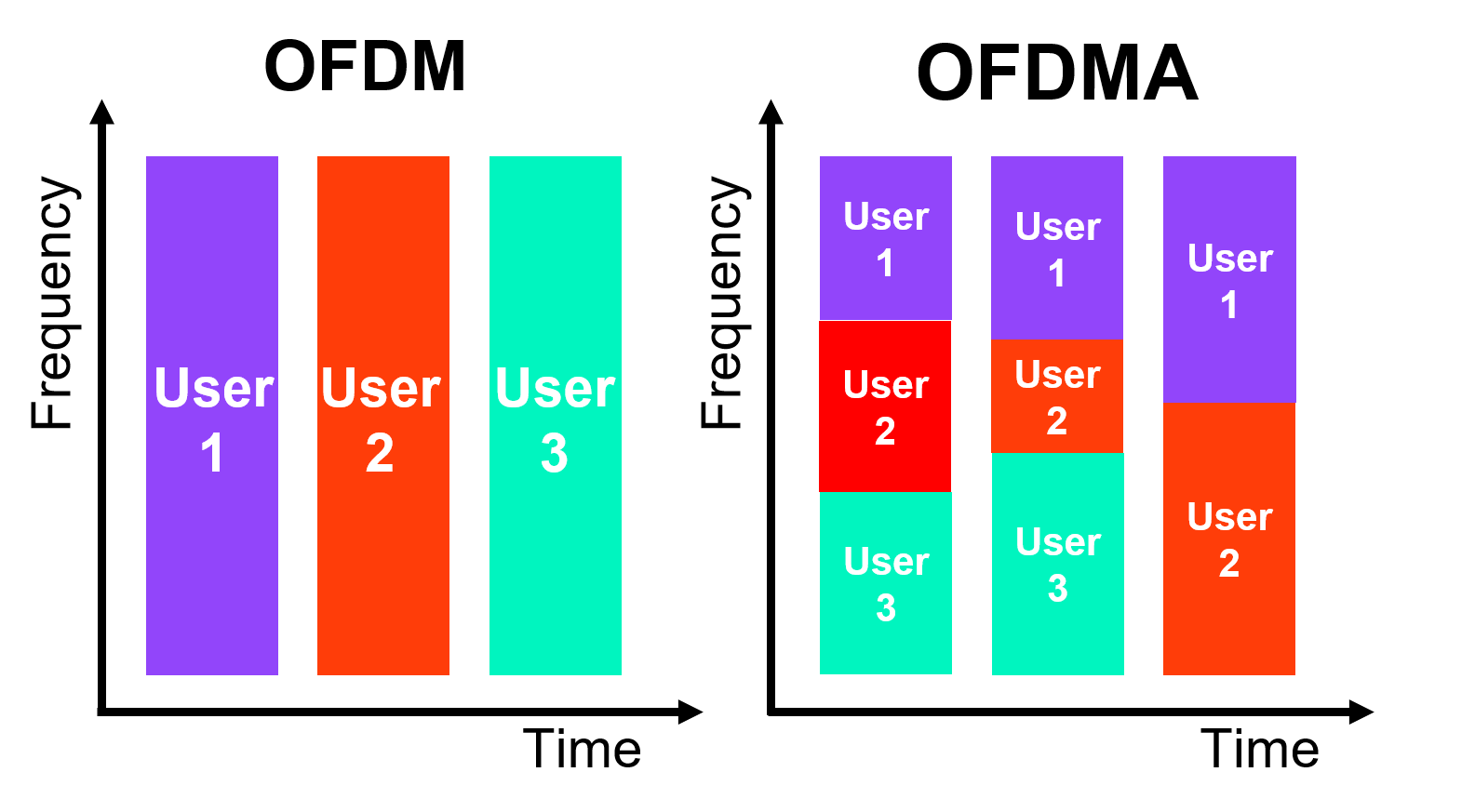}
    \caption{OFDM vs OFDMA Technology. \textmd{OFDMA enables multiple users to share the channel simultaneously, where the data of each user is transmitted on a subset of the channel bandwidth.}}
    \label{fig:ofdm_ofdma}
\end{figure}

\noindent\textbf{1) OFDMA:}
OFDMA is a new wireless technology introduced in 802.11ax (WiFi 6) and inherited in 802.11be (WiFi 7). It enables a WiFI AP to simultaneously communicate with multiple user devices by partitioning the channel bandwidth into multiple small trunks known as Resource Units (RU) and allocating different RUs to different users. Figure \ref{fig:ofdm_ofdma} illustrates the concept of OFDMA in comparison with Orthogonal Frequency Division Multiplexing (OFDM) which is the key modulation technology in the past few generations of WiFi. OFDM divides the WiFi channel bandwidth into multiple subcarriers and modulates data symbols on the subcarriers. It has the advantages of high spectral efficiency and robustness to channel impairments. However, OFDM is not designed for optimal resource allocation for multiple users. As illustrated in figure \ref{fig:ofdm_ofdma}, each OFDM packet only carries data for a single user, which causes insufficient spectrum usage and extra packet overhead. OFDMA solves this problem by allowing multiple users' data to be encapsulated in a single packet occupying different subsets of subcarriers. In this way, the spectral resources can be allocated dynamically according to the specific demands of user devices. OFDMA is anticipated to significantly reduce the communication latency in a dense deployment environment~\cite{qualcomm_white_paper}.

\begin{figure}[t!]
    \centering
    \subfloat[]{
        \includegraphics[width=2.4in]{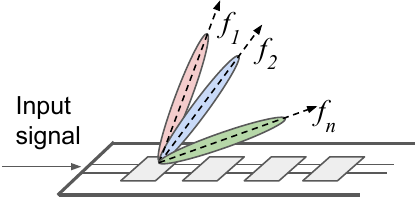}
        \label{fsa_concept}
    }
    \hfill
    \subfloat[]{
        \includegraphics[width=2.4in]{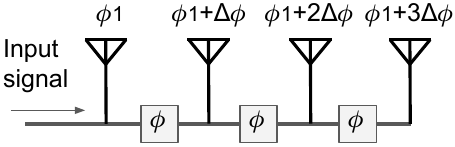}
        \label{fsa_principle}
    }
    \caption{FSA Principle. \textmd{(a) FSA is a passive structure which forms beams in different directions. The beam angle depends on the frequency of transmitted or received signal. (b) When the signal enters into the FSA structure, the signal experiences a linearly increasing phase shift at each antenna element, which combines the signal power constructively forming a beam. Specifically, the phase shift, which dictates the beam's direction, varies according to the signal frequency.}}
    \label{fig:fsa_principle}
\end{figure}

\vspace{0.1in}
\noindent\textbf{2) Frequency Scanning Antenna (FSA):}
FSA is a passive structure (similar to a patch antenna) that is mostly used in multi-dimensional radar image scanning~\cite{fsa_3dimage}. In contrast to traditional antennas, FSA receives and transmits a signal in a specific direction which depends on the frequency of the signal, as shown in Figure~\ref{fsa_concept}. The FSA structure consists of an array of elements connected by a transmission line as shown in Figure~\ref{fsa_principle}.
When a signal is fed to this structure, the signal experiences a different amount of phase shift at each radiating element. This natural phase shift depends on the frequency of the signal and causes the signals emitted at different antenna elements to combine constructively in a certain direction.
The emission direction can be adjusted by tuning the physical feed length, and element spacing, and adding additional band-pass filters between antenna elements~\cite{fackelmeier2010narrowband}. We build on this principle and design an FSA operating in the WiFi band. Our design optimizes the dimensions and field-of-view of the antenna with the constraint of limited bandwidth at WiFi bands. We will further discuss the implementation details of our FSA in Section~\ref{implementation}.




\section{Design Overview}
\name is a novel system that improves the communication and sensing capabilities of resource-constrained IoT devices. \name overcomes two main issues with today's IoT devices due to resource limitations. First, IoT devices have limited power budgets; therefore, they typically have a lower transmission power which limits its communication range and data rate. Second, IoT devices have limited hardware resources. In particular, most of the IoT devices today have only a single transceiver chain, which limits their capability to process signals for advanced capabilities such as localization. \name solves these two issues jointly with a novel smart antenna design.

Unlike traditional WiFi antennas, which distribute power uniformly in all directions and frequencies, \name's antenna passively focuses power for a particular frequency in a specific direction. We integrate this unique trait with the inherent features of WiFi protocols to enhance the communication and sensing capabilities of IoT devices.

We first show that \name can improve the communication performance of IoT devices by integrating the principle of FSA with OFDMA. In a dense WiFi network where IoT devices are typically distributed throughout the area, OFDMA allows multiple devices to communicate simultaneously with the AP by using different RUs. 
Traditionally, IoT devices use omnidirectional WiFi antennas, which transmit and receive all RUs with the same power density in all directions, resulting in sub-optimal power allocation while working with OFDMA. In contrast, \name concentrates antenna radiation power in the direction of the AP and in the RU allocated to the IoT device, which enhances the communication range and data rate.

We then show that \name can enable a single-chain IoT device to localize other WiFi devices. In particular, \name naturally enables the estimation of the signal's Angle-of-Arrival (AoA) by leveraging the correlation between frequency and spatial direction. By combining this direction information with the distance information obtained from packet round-trip time measurement, \name enables the localization of other WiFi devices in the vicinity. 

We organize the rest of the paper as follows. We first present \name's antenna design in Section \ref{fsa_design}. We then explain how to combine our antenna design with OFDMA to maximize the signal-to-noise ratio (SNR) in Section~\ref{sec:ofdma}. Next, we show how \name supports device localization in Section~\ref{sec:loc}.  Finally, we present the implementation details and evaluate results in Section~\ref{implementation} and~\ref{sec:eval}, respectively. 

\section{\name's Antenna Design}
\label{fsa_design}
\begin{figure}[t!]
    \centering
    \subfloat[Band-pass filter Design]{
        \includegraphics[width=0.55\linewidth]{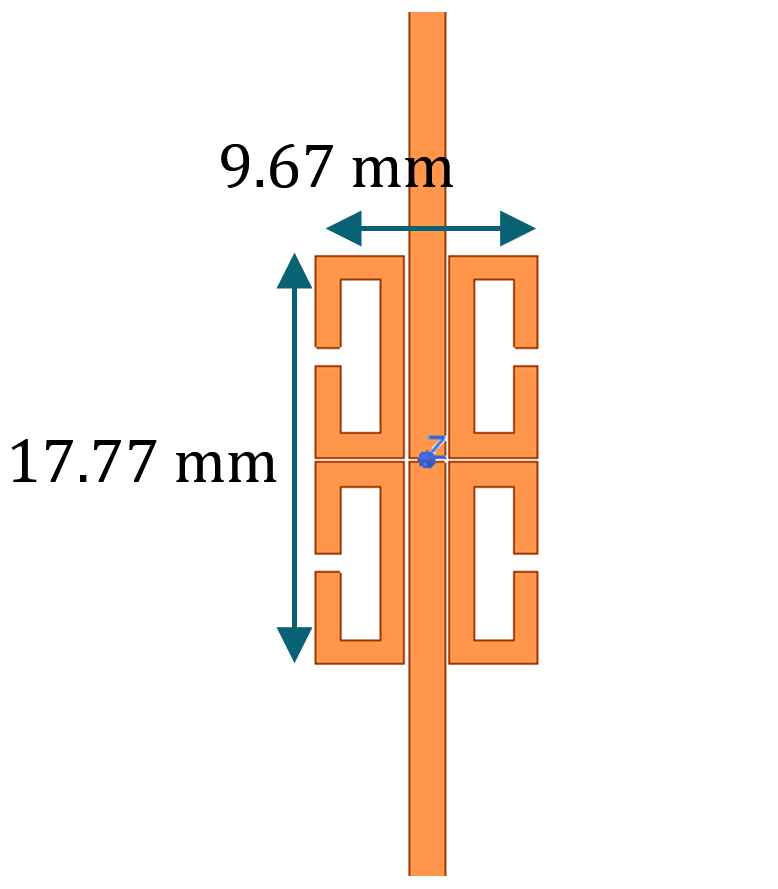}
        \label{fig:bpf}
    }
    \hfill
    \subfloat[T-Junction coupler design]{
        \includegraphics[width=0.8\linewidth]{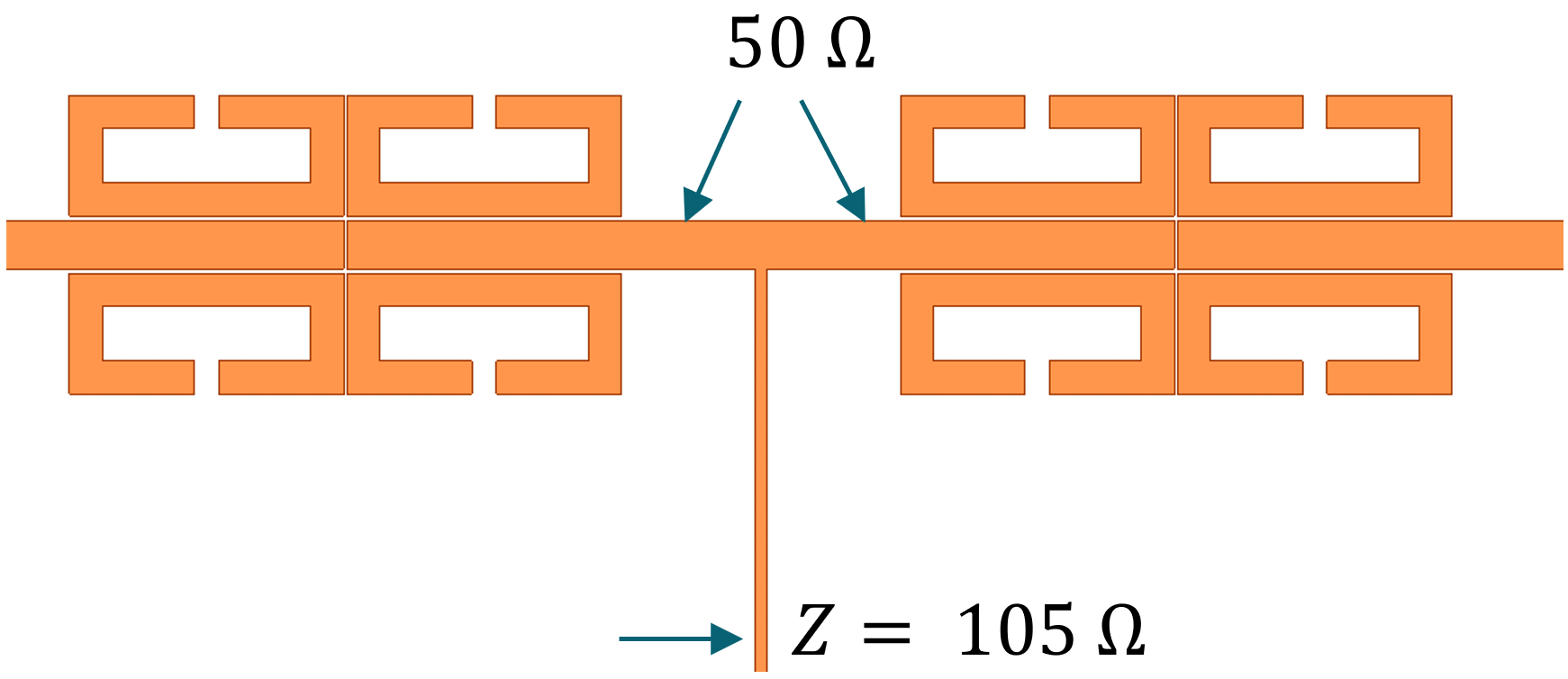}
        \label{fig:coupler}
    }
    \caption{\name's Antenna Components}
\end{figure}

We build on the principle of FSA to design \name's antenna. However, designing an FSA that operates effectively within the WiFi frequency bands presents several challenges:

The first challenge is to achieve a wide field-of-view angle with the limited bandwidth at WiFi bands. Nowadays, the common WiFi bands are at 2.4~GHz and 5.8~GHz. We designed our antenna to operate at the 5.8~GHz band, where there are 345~MHz of continuous bandwidth available. Unfortunately, this bandwidth is still considered limited to achieving a wide coverage angle compared to previous FSA designs~\cite{milback}, which typically require several GHz of bandwidth. The reason is that to achieve a large angular shift, the antenna structure needs to create a high phase variation for signals propagating in the structure. Conventionally, this is achieved using a long transmission line, which naturally causes more phase variation~\cite{nesic1995frequency}. However, due to the large wavelength at WiFi bands, transmission lines would have to be very long to create enough phase shifts, which could take up a lot of space. Some recent work~\cite{xu2019wide} proposed to use dispersive substrate integrated waveguides to implement passive phase shifters, which successfully increased the coverage angles without long transmission lines. However, the issue with this approach is the low energy efficiency, as almost half of power is dissipated within the antenna structure. Instead, our idea is to adopt bandpass filters as passive phase shifters, which create high phase variations with higher energy efficiency. Figure~\ref{fig:bpf} shows the structure of the band-pass filter. We implement the band-pass filters using split-ring-resonators~\cite{boskovic2017printed} which offers high phase variations with a compact size.  

The second challenge is to improve the energy efficiency of the antenna structure. The loss of energy is mainly caused by two factors: 1) the insertion loss of the band-pass filter; 2) the loss of the couplers that couple the signal power to the antennas. To reduce the loss of band-pass filters, we decrease the gap between the resonators to increase the coupling between them, which allows more power to be transferred. We use a simple T-junction for the coupler, as shown in Figure~\ref{fig:coupler}, which allows us to change the impedance of the middle transmission line (denoted as $Z$) to adjust the insertion loss and coupling factors. In particular, we designed the T-Junction with 105~$\Omega$ impedance to have both reasonable insertion loss and return loss. We use dipole antennas for the antenna elements to achieve higher gain and wider bandwidth. Different from previous FSA designs using dipole antennas, we place the dipole antennas perpendicularly on one side of the transmission line. In this way, we avoid creating extra beams in the bottom of the antenna to improve the energy efficiency of the antenna. Moreover, we place the dipole arrays at a tilted angle to reduce the coupling between the adjacent antenna elements, which further suppresses the undesired sidelobes.

\begin{figure}[t!]\centering
	\includegraphics[width=\linewidth]{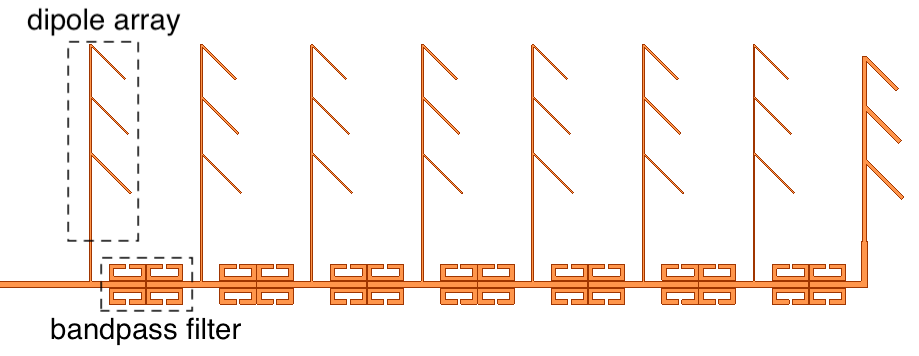}
	\caption{\name's antenna design}
	\label{fig:proposed_fsa}
\end{figure}
Finally, we want to make sure our antenna has a reasonable form factor to be installed on IoT devices. To achieve a compact design, we adopt a series feed structure for \name's antenna design, where we integrate the band-pass filters in between the antenna elements to increase the amount of phase shift, as shown in Figure~\ref{fig:proposed_fsa}. The structure consists of eight cascaded antenna elements connected by transmission lines with half-wavelength spacing. 

In terms of power, our antenna design is completely passive; therefore, incurs no additional power consumption to the IoT device. In terms of cost, the design costs a few dollars to fabricate since it just needs FR4 PCB, which can be further reduced with mass production.

\section{Enhacing IoT Communication using \name}
\label{sec:ofdma}
In this section, we show how \name integrates our smart antenna design with the OFDMA in the latest WiFi protocols to enhance the communication performance of dense IoT networks. 
As mentioned in Section~\ref{sec:background}, OFDMA allocates different resource units to different users, supporting simultaneous communication between AP and multiple IoT devices. However, this comes with the price of lower power allocated to each device, since a portion of the transmission power is no longer dedicated to the receiving device. Therefore, IoT devices communicating with the AP under OFMDA protocol suffer from lower communication range and data rate. \name solves this problem by amplifying the signal power within the RUs allocated to each IoT device using passive beamforming. 
Due to the frequency-dependent beamforming nature of FSA, signals received and transmitted in different directions will be amplified on different frequency ranges. \name utilizes this unique property of FSA combined with the RU allocation mechanism of OFDMA to enhance the communication performance for IoT devices.
As FSA is able to steer its beam in different directions, \name provides a relatively wide coverage angles while requiring no complex and power-hungry components such as a phased array. 

In the following paragraphs, we first discuss OFDMA's RU allocation mechanism that naturally aligns \name's beam direction to the direction of AP. Then we explain how \name benefits both downlink and uplink communication, respectively.

\vspace{0.05in}
\noindent\textbf{OFDMA RU Allocation:}
OFDMA's RU allocation protocol is not fixed by standard, a typical practice is to allocate RUs based on channel condition so that each device gets the RUs with their respective best channel to the AP~\cite{tutelian2021ieee}. Specifically, OFDMA AP first broadcasts pilot symbols to all the user devices engaged in the communication. Each user device estimates its channel and sends feedback to the AP regarding the channel condition. Based on the feedback from multiple users, the AP allocates RUs with the best channel condition to each user for communication. \name leverages this RU allocation mechanism to align its beam to the AP.

Specifically, when an IoT device is equipped with \name, depending on its direction with respect to the AP, some of the subcarriers of the pilot symbols are amplified by the beamforming gain. Therefore, in the feedback packets, the IoT device will inform the AP to use the RUs containing these strong subcarriers for communication. By selecting the appropriate RUs for each IoT device, the communication between AP and IoT devices benefits from the gain of \name's antenna.

\begin{figure}[t!]
    \centering
    \includegraphics[width=3.0in]{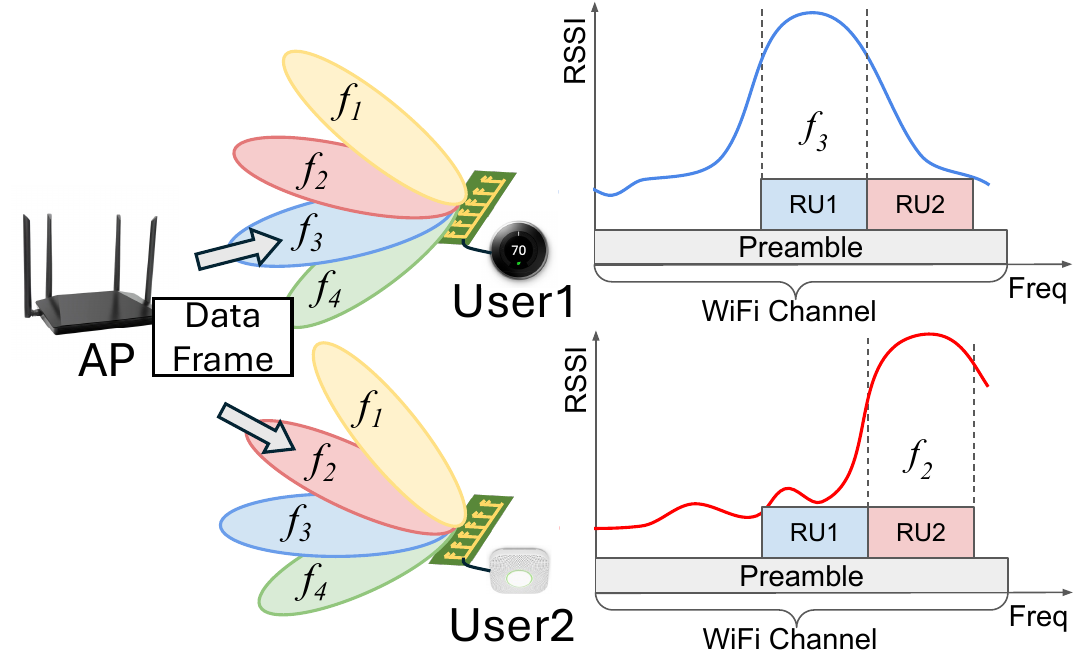}
    \caption{OFDMA Downlink. \textmd{When a user receives a signal from the direction of the AP, the signal power will be amplified in the RU assigned to that user in the frequency domain.}}
    \vspace{-0.1in}
    \label{ofdma_downlink_header}
\end{figure}
\vspace{-1pt}

\begin{figure}[t!]
    \centering
    \includegraphics[width=3.0in]{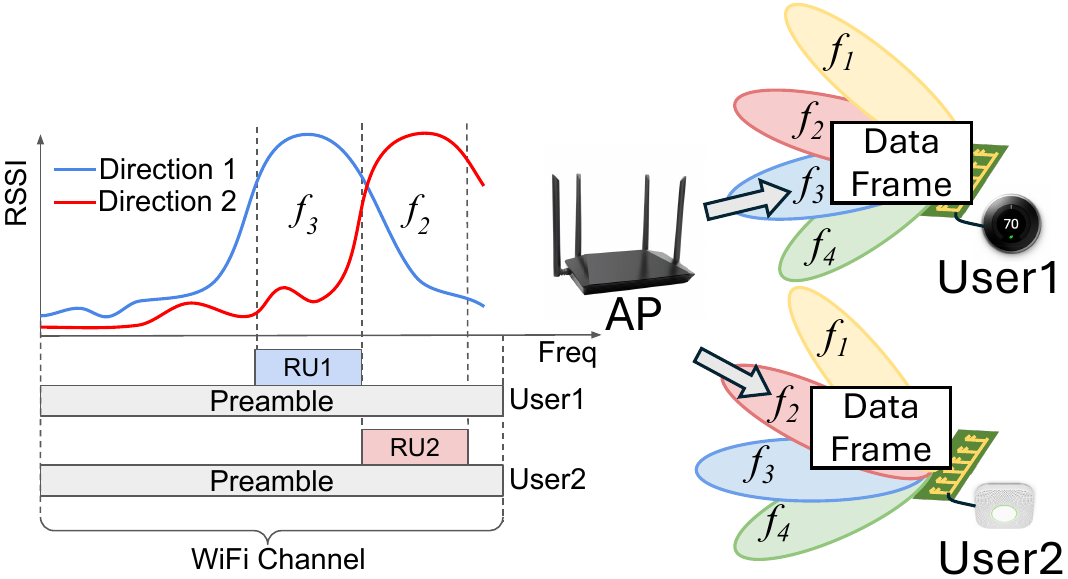}
    \caption{OFDMA Uplink. \textmd{The signal received by the AP from each user will be amplified in the RU assigned to that user.}}
    \vspace{-0.2in}
    \label{ofdma_uplink_header}
\end{figure}
\vspace{-1pt}

\vspace{0.05in}
\noindent\textbf{\name Downlink: } 
In OFDMA downlink communication, once the communication channel is reserved, the Access Point (AP) transmits data for multiple users within a single aggregated data frame. These RUs for different users share the same preamble which spans the entire channel bandwidth and carries common physical layer information to prepare the receiver for receiving subsequent data in the RU. 
One may be concerned that \name may potentially prevent the receiver from decoding the preamble, as it does not amplify the entire bandwidth equally. However, we find through experiment that this is not an issue, because 1) \name can still receive signals in other subcarriers than the corresponding beam, except with a lower gain; 2) the preambles are typically modulated using the lowest MCS scheme such as BPSK, which is more tolerable to lower SNR.
On the other hand, the data symbols contained in the RUs typically use higher order modulation schemes, which benefits from higher SNR.

Figure~\ref{ofdma_downlink_header} illustrates a scenario where two user devices communicate with the AP using different RUs. User~1 is allocated with RU~1, which corresponds to the frequency range around $f_{3}$. 
The signal received by User~1 from the AP will be amplified in the frequency range of $f_{3}$, as the beam for frequency $f_{3}$ is pointing in the direction of the AP.
This increases the received signal power for RU~1 which contains the data for User~1; thus boosting the SNR.
In the meanwhile, User~2, located in a different direction, is allocated with RU~2 in the frequency range of $f_{2}$. \name also amplifies the received signal power for RU~2 which contains the data for User~2, as the beam for frequency $f_{2}$ is pointing in the direction of the AP.

\vspace{0.1in}
\noindent\textbf{\name Uplink: } 
In OFDMA uplink, the AP first transmits a trigger frame to all the users to solicit the uplink transmission process. The users then transmit simultaneously on their allocated RUs in response to the trigger frame.
Each user transmits a separate data frame containing data in its allocated RU subcarriers, as illustrated in Figure~\ref{ofdma_uplink_header}.
Note that \name's antenna operates in a reciprocal manner. Therefore, each user's RUs are amplified in their corresponding beam directions. For example, in Figure~\ref{ofdma_uplink_header}, User~1 is located in Direction~1, and User~2 is located in Direction~2.
The frame transmitted by User~1 will be amplified in the frequency range of $f_{3}$. This is because signals within the $f_{3}$ frequency range add up constructively when it is transmitted from User~1 to the AP's direction. 
This increases the SNR of the data transmitted in RU~1.
Similarly, the frame transmitted by User~2 will be amplified in the frequency range of $f_{2}$, which increases the SNR of User~2's data transmitted in RU~2. 

In summary, \name enhances the OFDMA communication SNR in both downlink and uplink for resource-constrained IoT devices. This enhancement is achieved naturally by allocating the RUs for user devices corresponding to their beam directions.

\section{Enhancing WiFi Localization using \name}
\label{sec:loc}
In addition to having better communication range and data rate, IoT devices equipped with \name is also able to estimate the locations of other WiFi devices in the vicinity. Compared to conventional AP-based localization, enabling IoT devices to directly localize other devices has the advantage of low deployment costs. For scenarios that already contain many IoT devices, \name can be easily deployed by replacing the antenna. Even for environments without existing infrastructure, deploying an IoT device has a much lower cost than deploying a new AP to enable localization. Moreover, \name can localize devices no matter whether it is connected to the same network, which can enable a great variety of new applications. For instance, in environments where most of the devices are guest devices, such as a shopping mall, the localization system can work without requiring users to access the network. For security and surveillance systems, this can enable precise location tracking of assets and individuals, in case of unauthorized access or theft.

To localize a target device, two key pieces of information need to be measured: the distance to the target device and the relative angle to the target device.
Note that due to the cost limitation, most IoT devices today only have a single transceiver chain; therefore, they are not able to measure the AoA using traditional methods that require multiple chains. 
In this section, we show how \name measures distance and angle using a single chain.

\subsection{Measuring the AoA of Any WiFi Device}
\label{sec:localization_aoa}

First, we discuss how \name measures the AoA of any WiFi devices in the vicinity by leveraging the special property of FSA. As mentioned in Section~\ref{sec:background}, FSA creates beams of different frequencies in different directions, which produces a one-to-one mapping relation between the beam angle and frequency. 
For an IoT device equipped with \name, the received signal power is selectively amplified in the frequency range of the FSA beam in that direction.
Therefore, by analyzing the power distribution in the frequency domain, we can estimate the AoA of a WiFi packet.
To analyze the received signal power distribution in the frequency domain, \name analyzes the fine-grained Channel State Information (CSI) of the received WiFi frame. CSI characterizes the power distribution across the different subcarriers of a WiFi channel. This provides fine-grained information for AoA measurement.
Furthermore, CSI is embedded in the physical layer (PHY) header, meaning that \name could sniff the CSI information of devices in the area even if the WiFi packets are encrypted. 

Despite the advantages of CSI measurement, it is not supported by many IoT devices due to hardware constraints. To extend our solution, we utilize an inherent trait of the WiFi protocol which is supported by all WiFi devices -- WiFi probing.
Unlike regular data traffic packets, WiFi probe request packets are broadcasted by each WiFi device in every frequency channel periodically, to discover available networks in its communication range. Interestingly, we find that probe request packets are always transmitted by a WiFi device, regardless of whether the device is connected to a network or not. This is because connected devices also want to update their list of available access points in case of handover. 
As the probe request packet is transmitted on every 20 MHz frequency channel, \name is able to measure the AoA based on which channel has the highest Received Signal Strength Indicator (RSSI).
WiFi probing provides a natural tool for \name to measure the AoA of any WiFi device. Specifically, the IoT device can passively sniff the probe request packet of each target device broadcasting on each frequency channel. 


\begin{figure}[t!]
   \centering
   \includegraphics[width = 3.2in]{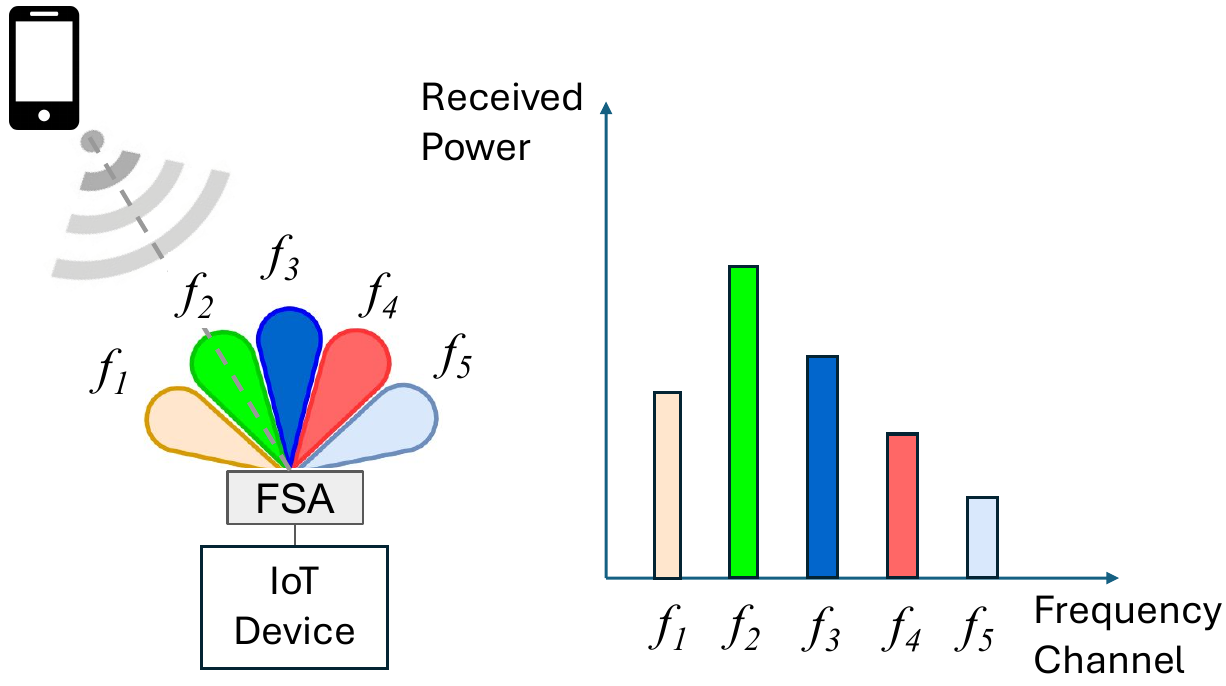}
   \caption{Measuring AoA using \name. \textmd{As WiFi devices periodically broadcast probe signals on different frequency channels, the localization device measures the RSSI profile of the received signal on each channel. The one-to-one mapping between the beam direction and frequency of FSA enables our system to interpolate the AoA of any WiFi device.}}
   \label{aoa_fsa}
\end{figure}

Figure~\ref{aoa_fsa} shows a simple example of how \name works. In this example, an IoT device is trying to measure the direction of a target device (a cell phone). For simplicity, we show an FSA with five directional beams corresponding to five channel frequencies ($f_{1}$ to $f_{5}$). The FSA is attached to the IoT device as the surrogate of its original antenna. As mentioned, the cell phone will periodically broadcast probe request packets in each frequency channel, the IoT device then simply measures the received signal power (RSSI) of the probe request packets. The right plot shows the received signal power the IoT device measures on each frequency channel. As illustrated, channel $f_2$ has the highest received power, since the corresponding beam direction aligns with the direction of the target device. Hence, by comparing the received signal power on each channel, the IoT device can estimate the signal DoA. 


\subsection{Measuring the Distance to any WiFi Device}
\label{sec:localization_dist}

\name measures the distance to the target device by measuring the time interval between the transmission of a data packet and the reception of its acknowledgment. This time interval includes the round-trip traveling time during which the packet is in the air and a Short Interframe Space (SIFS) that is fixed by standard. One can calculate the distance by removing the SIFS and multiplying half of the remaining time by the speed of light~\cite{WiFi-RTT}. This capability is introduced by the IEEE 802.11-2016 standard~\cite{IEEE802.11-2016} and has been supported in many WiFi devices. However, the standard requires a direct connection to be established between the measuring device and the target device, which typically appears between an AP and its client. For an IoT device to measure the distance to any other WiFi device, we need to trigger direct communication between the IoT device and its target. To address this challenge, we utilize a recently discovered phenomenon in today's WiFi devices. Specifically, researchers found that the PHY layer ACK frames are sent before validating whether the packet source~\cite{abedi2023wifi}, which implies that any WiFi device will send an ACK frame in response to a data frame (with the correct MAC address), even though the data frame may be fake and come from outside of the network. We utilize this property to trigger direct communication between the IoT device and any other WiFi device without establishing a connection link. Specifically, we let the IoT device first sniff the MAC address of a target device and forge a fake packet with the destination address set as the target MAC address. As the IoT sends this fake packet to the target device, an ACK frame will be triggered. The IoT device then can estimate the distance by the time difference between the transmission of the fake frame and the reception of ACK.

\section{Implementation}
\label{implementation}

\begin{figure}[t!]
	\centering
	\includegraphics[width=2.2in]{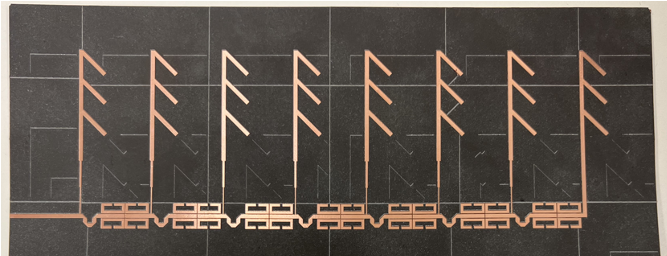}
	\caption{\name FSA Implementation. \textmd{It has a single port that can be easily connected to IoT devices instead of their current antenna. The design is also paper-thin and small. Hence, it can be placed on a wall, ceiling, or on an IoT box itself.}}
	\label{fsa_implementation}
\end{figure}                         
                     
We have designed and fabricated \name's FSA, as shown in Figure~\ref{fsa_implementation}. We design and simulate the antenna using CST~\cite{CST} software and fabricate it on RT/Duroid 5880 substrate. The antenna has a dimension of $22\times10$ $cm^2$ and a thickness of $0.5mm$.
It comprises a series of three-element dipole arrays, each with a $7.5$ dB gain. The first seven dipole arrays are designed to match the $105~\Omega$ T-junction couplers, and the last element matches $50~\Omega$. This antenna is optimized to operate at the 5.8~GHz ISM band, specifically within the frequency range of 5.5~GHz to 5.825~GHz. This covers 17 20~MHz channels over the 325~MHz bandwidth.
The structure is completely passive, with no active component. We use an SMA connector for the RF port of the antenna, which makes it plug-and-play with existing IoT devices.
For our experiments, we used Zimaboards~\cite{zimaboard}, each with an AX210 Network Interface Card (NIC) to emulate the IoT devices. We install Ubuntu 20.04 on each of the Zimaboards and use PicoScenes~\cite{jiang2021eliminating} for RSSI and CSI measurements. This gives us more flexibility to perform measurements and process data. However, the developed algorithms can be easily adapted to off-the-shelf IoT devices, such as ESP32. We use a TP-Link AX3000 Dual-band WiFi~6 router as the AP for experiments involving IoT communication with OFDMA protocols. 

\section{Evaluation}
\label{sec:eval}

In this section, we present the performance of \name. As a micro-benchmark, we first evaluate the performance of our fabricated FSA in terms of its antenna gain and direction-frequency correlation. Then, we assess the \name's effectiveness in extending the range and data rate of OFDMA communication. Finally, we provide evaluation results for device localization using \name, including the AoA and ToF measurement accuracy. 

\subsection{\name's Antenna Performance}
\begin{figure}[t!]\centering
	\includegraphics[width=0.82\linewidth]{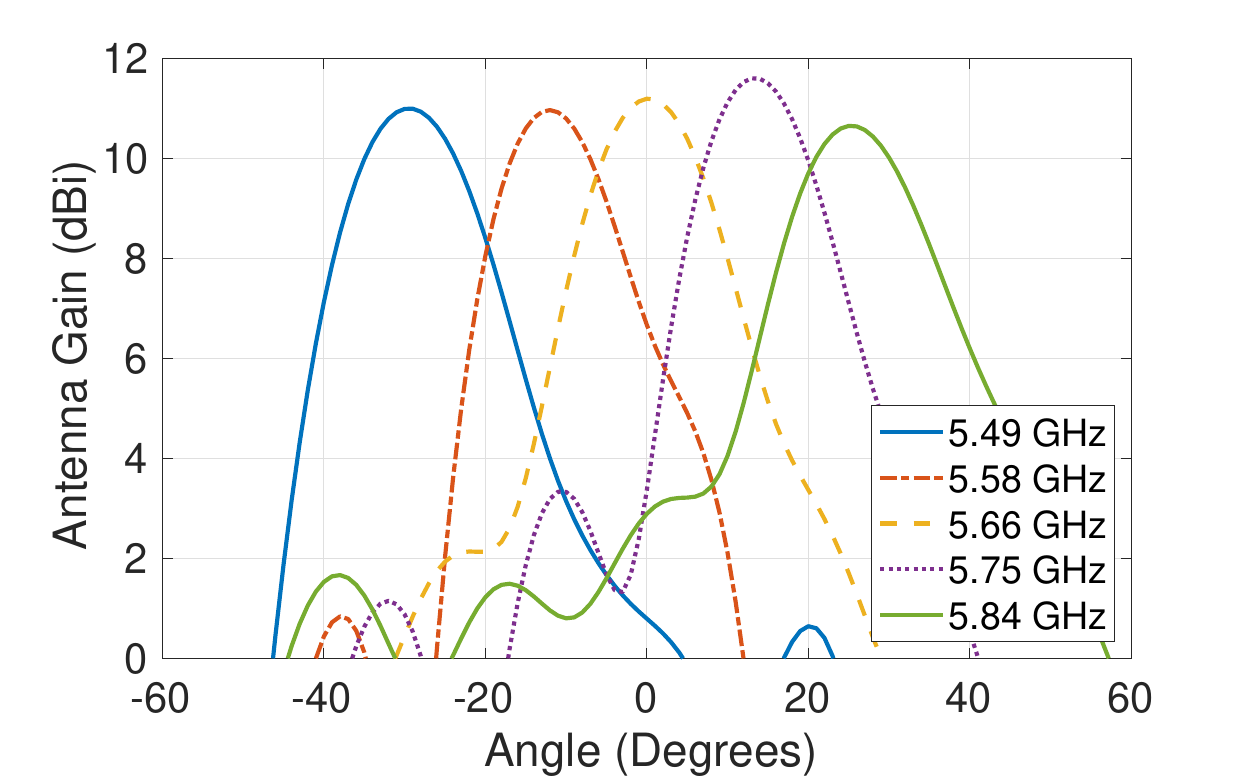}
	\caption{FSA Radiation Pattern. \textmd{We show the beam pattern for five discrete frequencies. Note that the beam direction steers continuously with frequency.}}
	\label{fig:fsa_result}
\end{figure}

\begin{figure}[t!]
    \centering
    \includegraphics[width=2.7in]{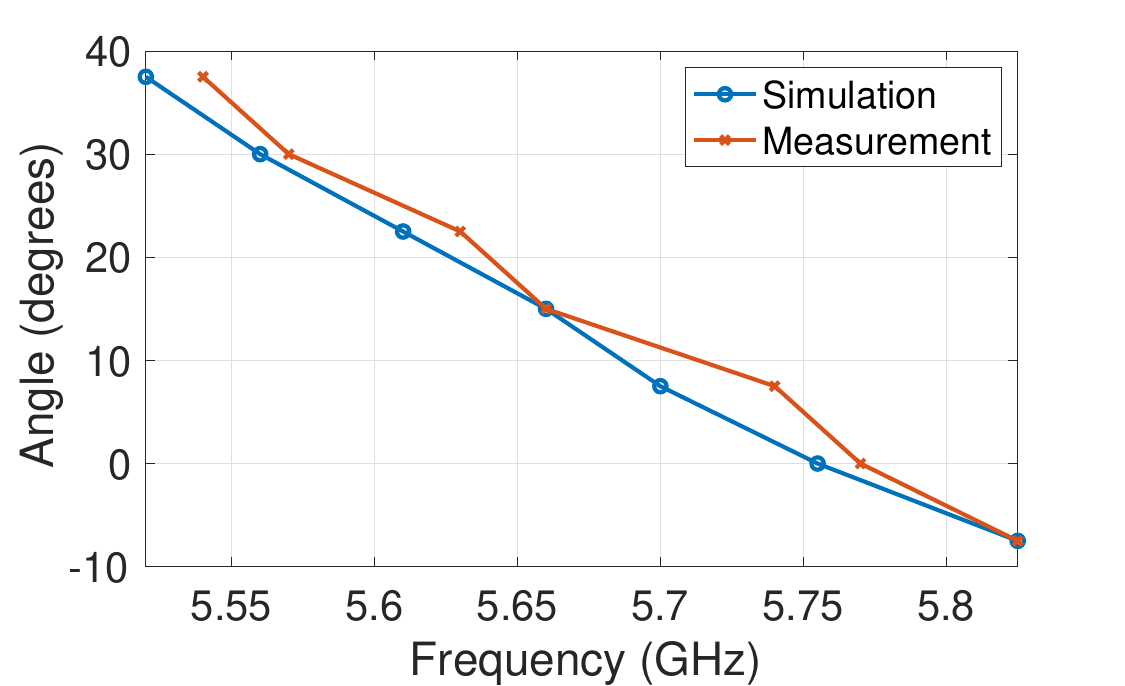}
    \caption{FSA Beam Angle Verses Frequency. \textmd{The FSA beam angle increases linearly with the increase in signal frequency. Our simulation results closely align with our real-world measurement results.}}
    \label{fsa_angle_vs_freq}
\end{figure} 

\begin{figure*}[t!]
    \centering
    \subfloat[0 degree]{
        \includegraphics[width=0.31\linewidth]{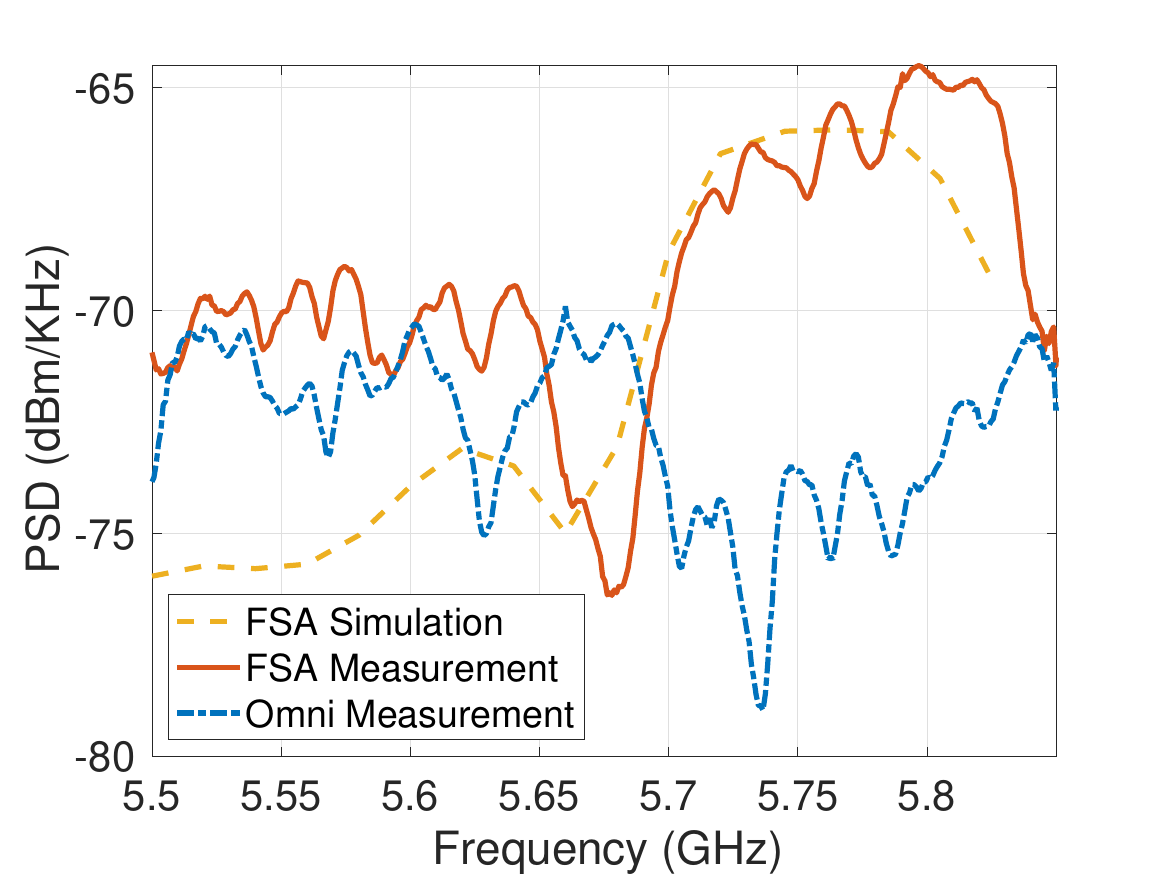}
    }
    \hfill
    \subfloat[15 degree]{
        \includegraphics[width=0.31\linewidth]{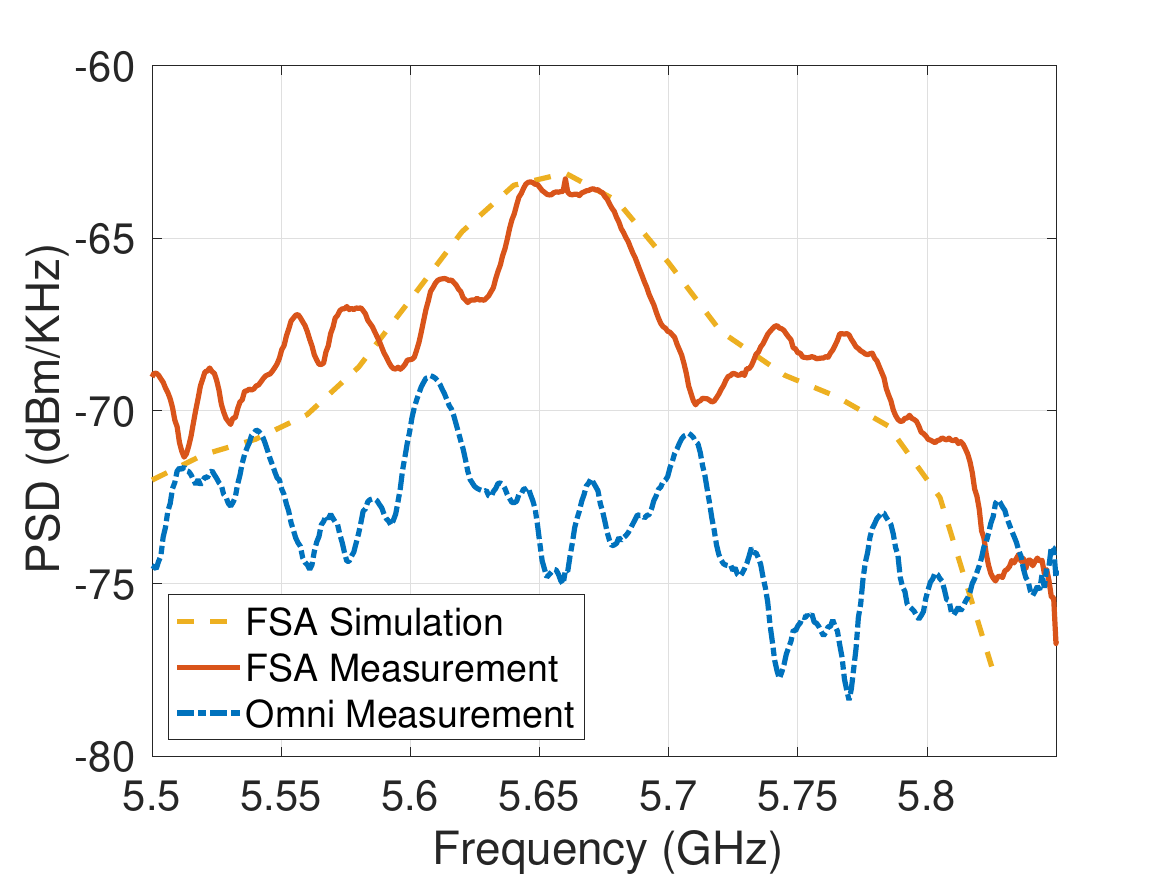}
    }
    \hfill
    \subfloat[30 degree]{
        \includegraphics[width=0.31\linewidth]{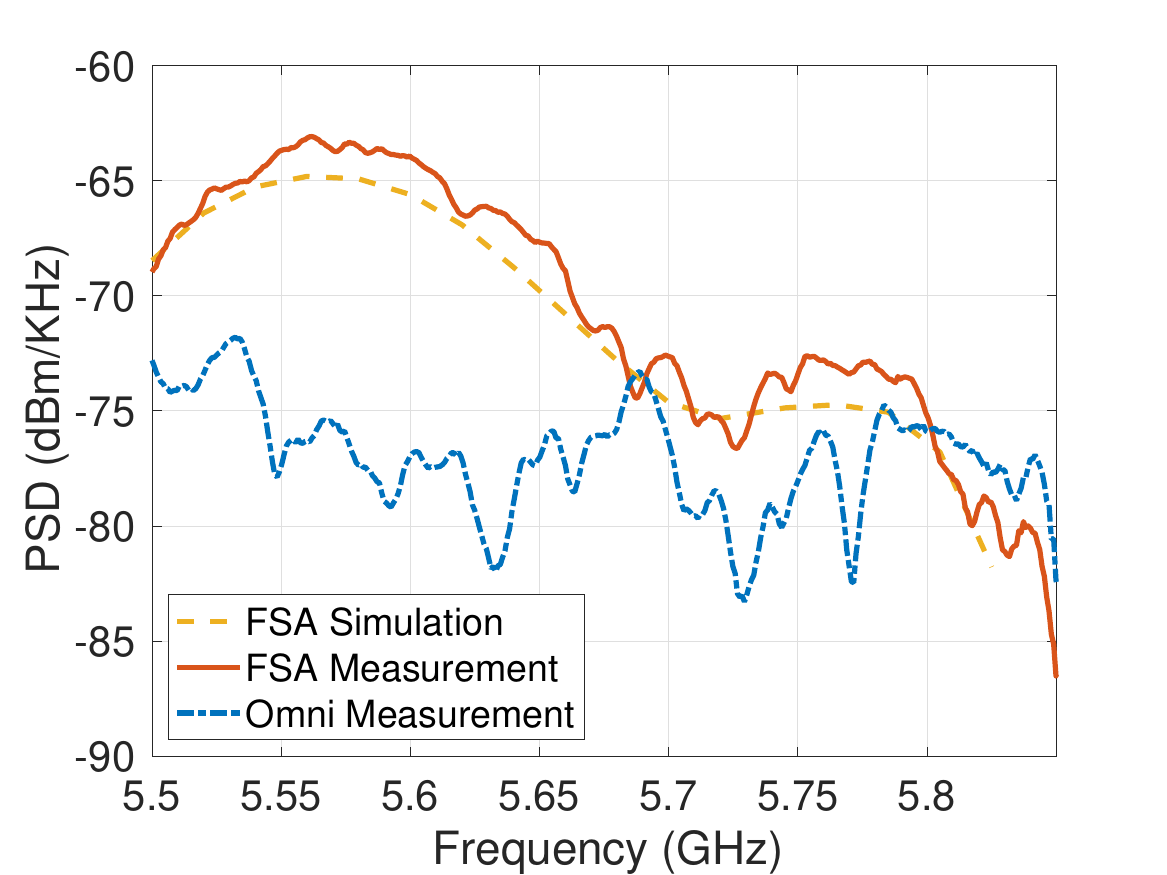}
    }
    \caption{FSA Power Distribution Across Frequency at Different Angles. \textmd{This shows the received signal power of the FSA and omnidirectional antenna across the frequency domain when the transmitter is placed in different directions. We also show the simulation result of the FSA radiation pattern in the frequency domain with PSD value adjusted to the measurement results.}}
    \label{fig:fsa_profile}
\end{figure*}

Figure~\ref{fig:fsa_result} shows a simulation result of the radiation pattern of the FSA. The x-axis is the azimuth angle with respect to the normal direction of the transmission line, and the y-axis is the antenna's gain in dBi. We plot the beam pattern for five discrete frequencies for the simplicity of illustration. It is worth noting that in reality, FSA steers its beam continuously with the change of frequency. The beam direction (indicated by the peak gain) steers with the increase in frequency. Our FSA achieves over 10 dBi gain over the whole bandwidth. The total steering angle is over 50 degrees. This coverage angle is sufficient,  considering that typical IoT devices are usually mounted on the walls or corners of a room. 

We then conduct real-world experiments to validate the simulation results. We connect the FSA to a signal analyzer and use a signal generator, placed 1.5~m away from the analyzer, to transmit a wideband signal spanning the entire operational frequency range of the FSA (5.5 GHz to 5.825 GHz). We measure the power spectral density (PSD) of the received signal while placing the transmitter in various directions.
Figure~\ref{fsa_angle_vs_freq} shows the correlation between the beam direction and signal frequency based on both simulation and real-world measurements. The results exhibit that there is a linear correlation between the FSA's beam direction and the signal frequency. 

We further show the measured power spectral density (PSD) of the FSA in Figure~\ref{fig:fsa_profile}. 
For comparison, we place a regular dipole WiFi antenna with 3~dBi gain (omnidirectional) at the same location as the FSA to measure the received signal power.
As shown in the figures, the signal power density increases as the frequency aligns with the beam direction. The variation of PSD matches closely with the simulation beam pattern of the FSA for the corresponding angle. In contrast, when the receiver uses the omnidirectional antenna, the power density is more uniformly distributed across the frequency domain. Overall, we show that the fabricated antenna has the expected beam pattern we designed. Finally, we swap the antenna of the transmitter and receiver and run the same experiment. The results show a similar pattern in the received power spectrum. This verifies that our FSA design is reciprocal; therefore, can be used for both uplink and downlink communications.


\subsection{Communication Performance}

\begin{figure}[t!]
    \centering
    \includegraphics[width=2.2in]{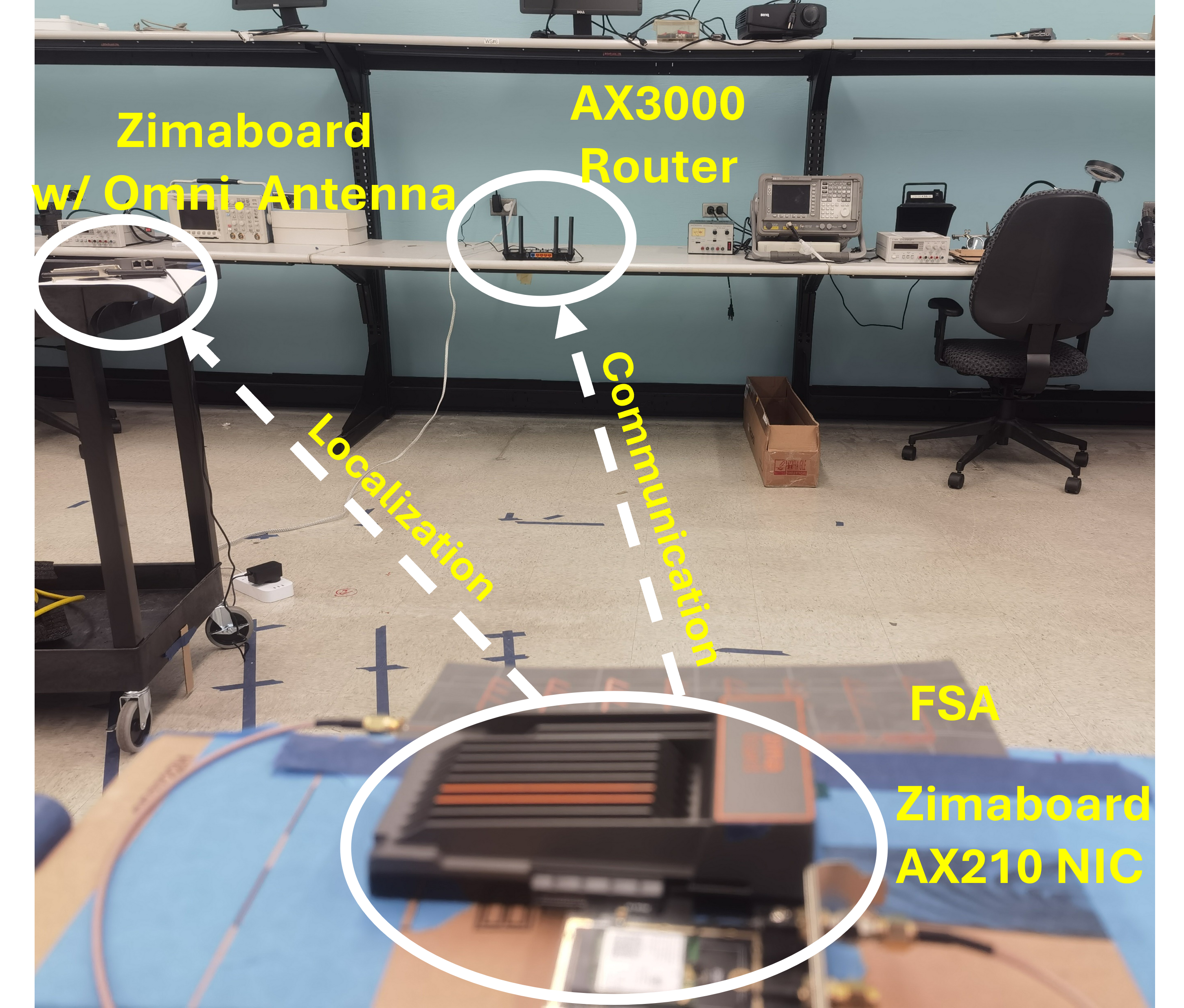}
    \caption[]{Indoor Experiment Setup}  
    \label{office_setup}
\end{figure} 


\begin{figure*}[t!]
    \centering
    \subfloat[SNR Improvement]{
        \includegraphics[width=0.31\linewidth]{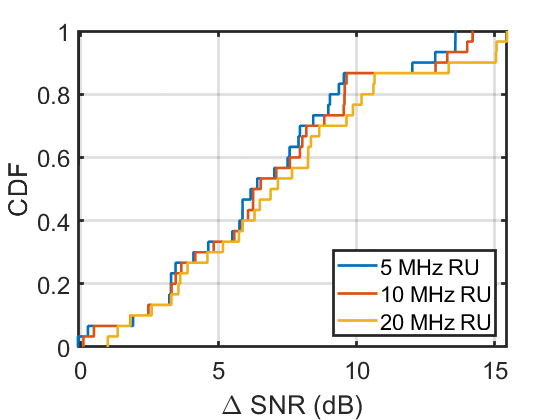}
        \label{measurement_snr_increase_cdf}
    }
    \hfill
    \subfloat[Data Rate Improvement]{
        \includegraphics[width=0.31\linewidth]{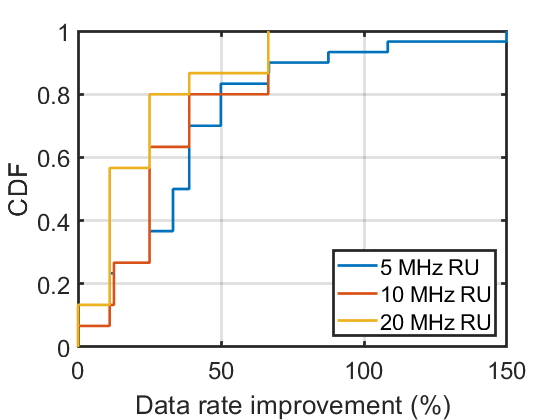}
        \label{measurement_rate_increase_cdf}
    }
    \hfill
    \subfloat[Distance Improvement]{
        \includegraphics[width=0.31\linewidth]{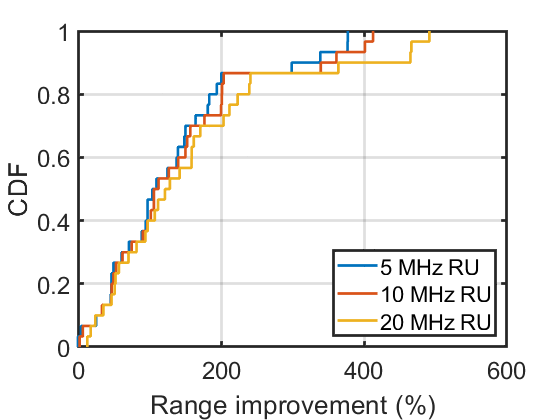}
        \label{measurement_distance_increase_cdf}
    }
    \caption{Performance Improvement Measurement. \textmd{The communication performance improvement of \name compared to omnidirectional antenna when the user device is communicating using 5 MHz, 10 MHz, and 20 MHz RUs.}}
    \label{measurement_indoor}
\end{figure*}

\subsubsection{Measurement Results}
We first evaluated the performance of our system by performing real-world measurements in different indoor environments: an office with lots of desks and cubicles, as well as a lab space. We collected measurements at 15 different locations.
Our setup is shown in Figure~\ref{office_setup}. We set a TP-Link AX3000 Dual-band WiFi~6 router at a fixed location to work as an AP and put multiple Zimaboards equipped with \name at random locations in the environment to emulate the IoT devices. We let the communication devices transmit regular data packets with a bandwidth of 160~MHz and collect the CSI of the received signal using the Zimaboards at each location. Each Zimaboard is connected to an AX210 NIC where we terminate one of the two RF ports and attach the other port to our FSA to emulate a single-chain IoT device. 
As a baseline, we connect an omnidirectional dipole antenna with 3~dBi gain to the same port instead of the FSA. We use SNR as a metric for comparison.

We assume OFDMA divides the 160~MHz channel bandwidth uniformly into $m$ RUs and allocates the RU with the best channel quality to each IoT device. We consider different RU bandwidths from 5~MHz to 20~MHz. We use the following method to calculate SNR improvement for each RU when using FSA with respect to omnidirectional antenna:

Let $P_{\text{A}, i}$ denote the received signal power of \(i\)-th RU when using antenna type \text{A} at the IoT device, where $A\in \{FSA, Omni\}$, $i\in [1, m]$ and \textit{Omni} stands for an omnidirectional antenna. Then the SNR improvement is formulated as follows:

\begin{equation}
\Delta \text{SNR} (\text{dB}) = \max_{i=1}^{m} P_{\text{FSA}, i} (\text{dBm}) - \max_{j=1}^{m} P_{\text{Omni}, j} (\text{dBm})
\end{equation}

Basically, we pick the best RU for each case and calculate their difference in dB.

We collect 150 packets in both uplink and downlink for each location and antenna type and compare the median values over all the packets for antenna type to calculate the SNR improvement.
Figure~\ref{measurement_snr_increase_cdf} shows the results for three different RU bandwidths. We can see that our FSA achieves much higher SNR compared to the omnidirectional antenna. More specifically, 50\% of the time using FSA is able to achieve more than 7 dB higher SNR compared to using an omnidirectional antenna.

The enhancement of SNR implies that IoT devices now can use higher modulation and coding schemes for communication, which increases the data rate.
Figure~\ref{measurement_rate_increase_cdf} shows the corresponding data rate improvement. For each location, we find the highest data rate achievable with the available SNR using the Modulation and Coding Scheme (MCS) table for 802.11ax~\cite{MCSIndex2023}. Then we calculate the relative improvement when using our FSA compared to an omnidirectional antenna. As shown in the figure, the highest data rate increases for evaluated RU bandwidths. 
 We are able to achieve an improvement of around 25\% (10th percentile: 11\%, 90th percentile: 80\%) for different RU sizes on average. For 5MHz RU, we achieve up to 150\% improvement.

We further evaluate the range improvement. For each measurement, we calculate the communication range for the lowest MCS scheme by extrapolating the measured SNR with the free space path loss (FSPL) formula~\cite{molisch2012wireless}. As shown in the figure, \name improves the communication range by 113\% (10th percentile: 34\%, 90th percentile: 388\%) on average for different RU sizes. It can achieve up to five times range improvement for 20MHz RU.

\subsubsection{Large-scale Simulation}

\begin{figure}[t!]
    \centering
    \subfloat[Indoor Simulation Setup]{
        \includegraphics[width=2.5in]{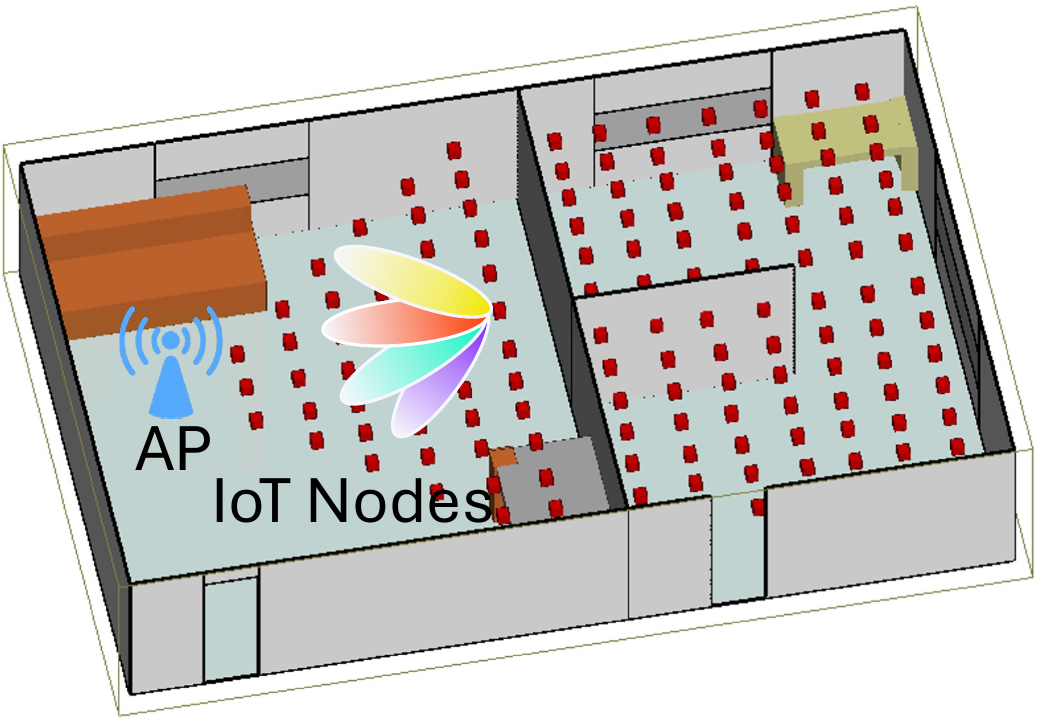}
        \label{simulation_indoor_setup}
    }
    \hfill
    \subfloat[SNR Increase]{
        \includegraphics[width=2.8in]{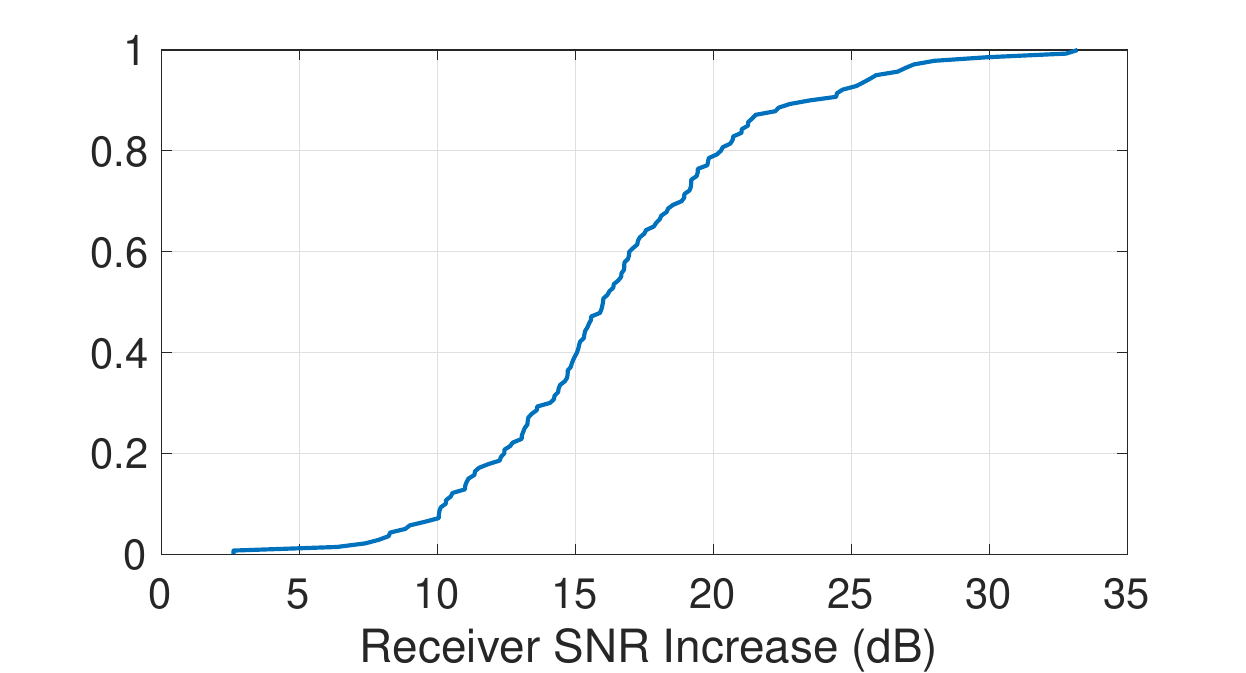}
        \label{simulation_snr_increase_cdf}
    }
    \caption{SNR Improvement Simulation. \textmd{(a) We placed the AP on one side of the building and uniformly distributed 140 IoT devices in different rooms of the building. We compared the performance of FSA and omnidirectional antenna on each IoT device. (b) The SNR improvement of FSA compared to omnidirectional antenna.}}
    \label{simulation_indoor}
\end{figure}

To further demonstrate the effectiveness of \name under more complex scenarios (with multipath reflections) and a large number of devices, we conduct a large-scale simulation. We use Remcom Wireless InSite~\cite{wirelessinsite} to simulate the signal propagation in an indoor environment, as shown in Figure~\ref{simulation_indoor_setup}. The environment is of dimension $16\times10\times3 m$ with various reflectors and absorbers such as walls and furnitures.
We set up a fixed access point in the environment and set 140 IoT devices (each equipped with an FSA) uniformly distributed in the environment. We use an ideal FSA with eight beams pointing in different directions.
Similar to our design, the total coverage of the FSA beams is 60 degrees, and each beam has a center gain of 15 dBi. We let each beam transmit a signal using one of the non-overlapping 20 MHz frequency blocks in the range of 5.50 GHz to 5.66 GHz (160 MHz channel).

For comparison, we run the same simulation except using an omnidirectional antenna on each of the IoT devices. For each of the 140 locations inside the rooms, we recorded the max SNR across all the 20 MHz subchannels for both antenna types. 
We then calculate the improvement in SNR when the IoT devices use FSA compared to omnidirectional antenna.

We plot the CDF of the SNR improvement in Figure~\ref{simulation_snr_increase_cdf}. 
As shown in the figure, FSA can achieve SNR improvement 100\% of the time. Moreover, FSA can achieve at least 17 dB of SNR improvement at 50\% of the time. 

\subsection{Localization Performance}

\vspace{0.1in}
\subsubsection{AoA Measurements}
\begin{figure}[t!]
    \centering
    \subfloat[CSI Measurement at 23 Degrees]{
        \includegraphics[width=0.8\columnwidth]{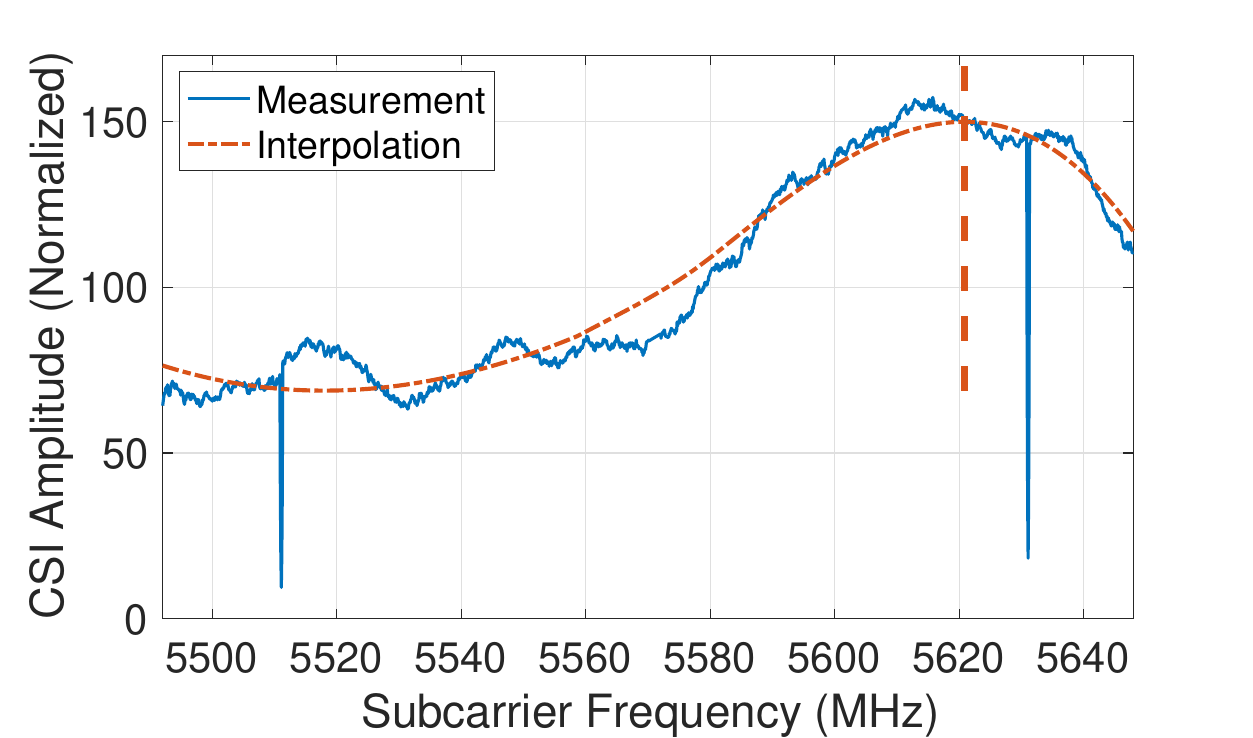}
        \label{fig:aoa_interpolation_23d}
    }
    \hfill
    \subfloat[AoA Estimation Accuracy]{
        \includegraphics[width=0.8\columnwidth]{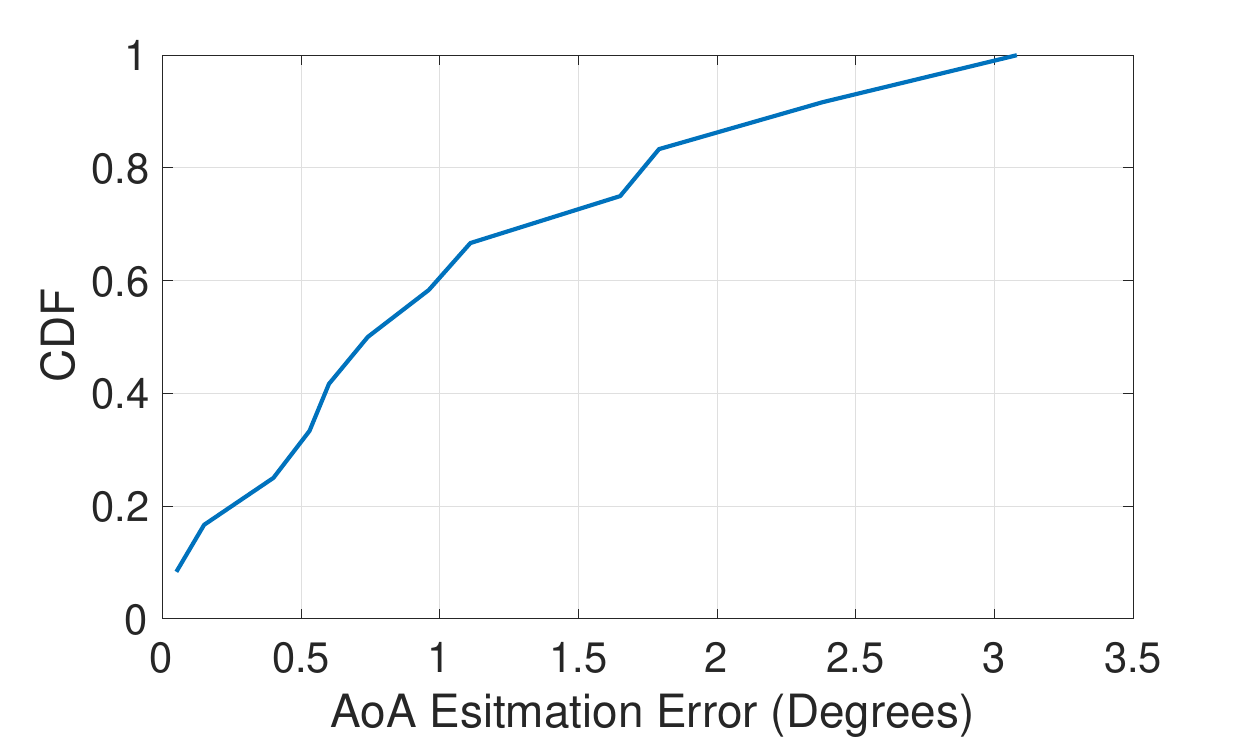}
        \label{fig:aoa_error_cdf}
    }
    \caption{\name AoA Estimation Performance.\textmd{(a) The normalized CSI amplitude measurement of the received WiFi packet, when the transmitter is placed at 23 degrees relative to the localization device. (b) AoA estimation accuracy using \name.}}
    \label{fig: aoa_estimation} 
\end{figure}

For our AoA measurement, we set up a Zimaboard as the localization device and connected it to the FSA for receiving signals.
We transmit 802.11ax packets with a center frequency of 5.57~GHz and 160~MHz bandwidth. We measure the ground truth with a protractor and a laser pointer. We pose the target device at 15 different locations with various distances and angles from the device performing localization.
To calculate the AoA, we take the CSI of the received signal and apply spline interpolation and Savitzky-Golay smoothing filter to reduce noise. We then identify the peak frequency and use the direction-frequency correlation as shown in Figure~\ref{fsa_angle_vs_freq} to find the corresponding AoA. Figure~\ref{fig:aoa_interpolation_23d} shows the CSI magnitude across the 160 MHz WiFi channel when the target device is placed at 23 degrees with respect to the localization device. We plot both the raw signal and the signal after interpolation and smoothing. From the figure, we estimate the peak frequency to be at 5.620~GHz, which corresponds to an estimated AoA of 23.6 degrees, resulting in only 0.6-degree estimation error.
We plot the CDF of the AoA estimation error in Figure~\ref{fig:aoa_error_cdf}. As is shown, \name can achieve an AoA estimation error of less than 1 degree 50 \% of the time, with a max estimation error of around 3.1 degrees.

\subsubsection{Distance Measurements}
We measure the distance between the target device and the device performing localization using round-trip time-of-flight, as mentioned in Section~\ref{sec:localization_dist}

\begin{figure}[t!]\centering
         \centering
         \includegraphics[width=0.7\linewidth]{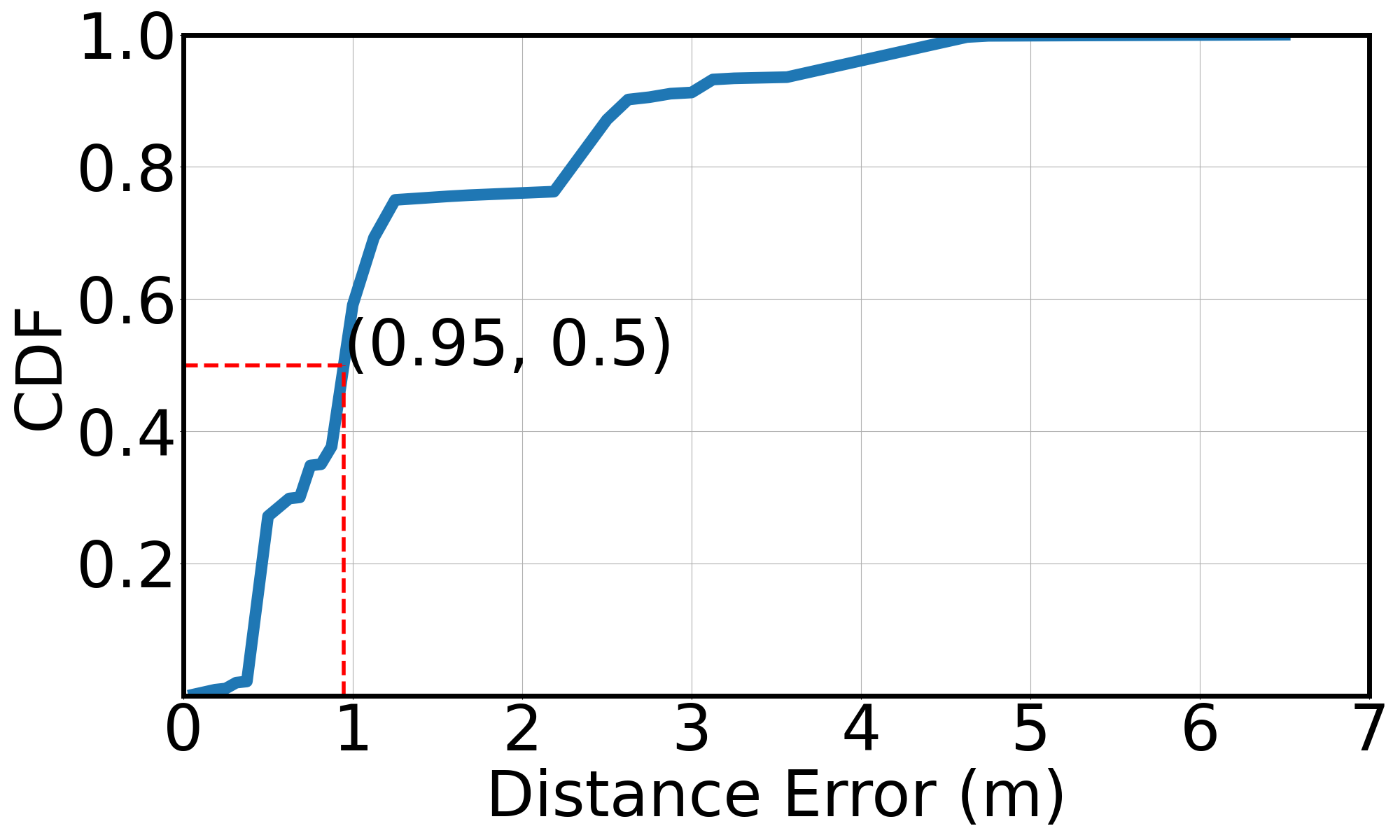}
         \caption{Distance Estimation Accuracy}
        \label{fig:dist_err_cdf}
\end{figure}

In this experiment, we place the target device at various distances from the localization device. The localization device sends 100 fake data packets per second and measures the ToF. Samples collected within 1 second (100 DATA-ACK pairs) are chosen as a sample group. The median value of the sample group is taken for the distance estimation. We collected 150 sample groups for each distance. We measure ground truth distance using a BOSCH GLM20 laser distance measure. Figure~\ref{fig:dist_err_cdf} shows the Cumulative Distribution Function (CDF) of the distance estimation error. We note that 80\% of the estimated distances have an error of fewer than 2.5m and the median error is 0.95m. It is worth mentioning that although past work has achieved higher accuracy in distance measurement, they require much more complex WiFi devices. In contrast, \name enables localization on low-power IoT devices that have only one transceiver chain. Moreover, \name's accuracy is sufficient for many WiFi localization applications. 

\section{Related Work}
\label{sec:related}

This paper is an extension of our previous workshop paper~\cite{wisight}. The workshop paper proposed an initial idea for bringing localization to IoT devices using FSA and provided some preliminary results. However, it did not address challenges in designing and fabricating an FSA for WiFi and did not present an evaluation of the end-to-end system. In contrast, here we address all these challenges, design and fabricate an FSA at WiFi frequency. We further propose a complete system that incorporates our FSA design with the existing WiFi protocols to improve both communication and sensing. Specifically, we integrate frequency scanning techniques with OFDMA to enhance the communication range and data rate leveraging both spectral and spatial resources of the wireless network. We also propose and evaluate a more accurate AoA measurement using the CSI across different subcarriers of the 802.11ax packet. Finally, we conducted real-world experiments to evaluate the performance of our end-to-end system. Below we address other related work.

\vspace{0.1in}
\noindent\textbf{Leaky-Wave Antenna (LWA) and Frequency Scanning Antenna (FSA)}
\name's antenna design builds on past work of Frequency Scanning Antenna (FSA) and Leaky-Wave Antenna (LWA). These antennas provide a directional radiation pattern that can be controlled by altering the frequency of the signal~\cite{lwa}. Unlike past work which uses them for direction finding at THz and mmWave bands~\cite{saeidi2021thz,ghasempour2020single,li20225g}, our design works at  WiFi band with limited bandwidth, thus facing many unique challenges. There are some past work on FSA operating at sub-6GHz bands~\cite{fsawifi_gil2021wi,fsawifi_paaso2017doa,fsawifi_poveda2020rssi,fsawifi_gil2023monopulse}. However, neither of them explore the integration of FSA with WiFi protocol and enabling more efficient OFDMA and localization. Moreover, in contrast to past work, we designed angled dipole arrays and integrated them into our FSA as its radiating elements. This enables much lower coupling between FSA elements, higher beam gain, and fewer side lobes which is necessary in integrating the FSA with OFDMA and localization algorithms. In summary, \name introduces a new WiFi antenna based on FSA and combines it with the latest WiFi medium access protocol to provide improvement in both communication and sensing, which has not been demonstrated in past work.

\vspace{0.1in}
\noindent\textbf{Integrated Sensing and Communication (ISAC):} 
Integrated Sensing and Communication (ISAC) is a new design paradigm for the next-generation wireless networks. The objective is to design wireless systems that are capable of both sensing the surroundings and communicating with networked devices to provide various services. Unlike many previous wireless sensing schemes that support sensing at the price of compromising communication performance, ISAC requires mutual improvement in both sensing and communication~\cite{liu2022integrated, isac-survey}. There have been some recent attempts to design ISAC systems at both signal and system level. At signal level, \cite{liu2018toward,liu2020joint} propose designs of special waveforms to enable communication on the existing wireless radars. \cite{sturm2011waveform} adapt OFDM waveform to facilitate sensing. At the system level, \cite{he2023sencom} extends WiFi CSI measurements to general WiFi packets and unifies the CSI interface for upper-layer sensing applications. \cite{milback} introduces an integrated sensing and communication system based on millimeter wave (mmWave) backscatter. In contrast, \name focuses on enabling ISAC for resource-constrained IoT devices. We introduce a passive beamforming network that can be easily integrated with low-power, low-cost IoT devices to enhance both communication and localization capability. \name can significantly improve the capability of the existing IoT infrastructures with very low extra deployment costs.
\section{Conclusion}
\label{sec:conclusion}

This paper presents \name, a system that seamlessly integrates smart antenna hardware with today’s WiFi protocol to enhance the communication and sensing
capabilities of resource-constrained IoT devices.
\name not only enhances the communication range and data rate significantly for dense IoT networks using OFDMA protocol; but also enables low-cost, low-power IoT devices with a single transceiver chain to localize other WiFi devices. We believe this work represents an important step towards the unified design of sensing and communication systems for future IoT networks.

\section*{Acknowledgments}
We thank all members of UCLA ICON lab for their comments and feedback, in particular Mohammad Mazaheri and Reza Rezvani for helping with the antenna fabrication. This work is partially funded by CISCO and UCLA.

\bibliographystyle{IEEEtran}
\bibliography{ref}

\begin{thebibliography}{10}
\providecommand{\url}[1]{#1}
\csname url@samestyle\endcsname
\providecommand{\newblock}{\relax}
\providecommand{\bibinfo}[2]{#2}
\providecommand{\BIBentrySTDinterwordspacing}{\spaceskip=0pt\relax}
\providecommand{\BIBentryALTinterwordstretchfactor}{4}
\providecommand{\BIBentryALTinterwordspacing}{\spaceskip=\fontdimen2\font plus
\BIBentryALTinterwordstretchfactor\fontdimen3\font minus \fontdimen4\font\relax}
\providecommand{\BIBforeignlanguage}[2]{{%
\expandafter\ifx\csname l@#1\endcsname\relax
\typeout{** WARNING: IEEEtran.bst: No hyphenation pattern has been}%
\typeout{** loaded for the language `#1'. Using the pattern for}%
\typeout{** the default language instead.}%
\else
\language=\csname l@#1\endcsname
\fi
#2}}
\providecommand{\BIBdecl}{\relax}
\BIBdecl

\bibitem{qualcomm_white_paper}
``The benefits of ofdma for wi-fi 6,'' \url{https://www.qualcomm.com/content/dam/qcomm-martech/dm-assets/documents/ofdma_white_paper.pdf}, accessed: 2023-09-16.

\bibitem{fsa_3dimage}
S.~Zhan-shan, R.~Ke, C.~Qiang, B.~Jia-jun, and F.~Yun-qi, ``3d radar imaging based on frequency-scanned antenna,'' \emph{IEICE Electronics Express}, vol.~14, no.~12, pp. 20\,170\,503--20\,170\,503, 2017.

\bibitem{fackelmeier2010narrowband}
A.~Fackelmeier and E.~Biebl, ``Narrowband frequency scanning array antenna at 5.8 ghz for short range imaging,'' in \emph{2010 IEEE MTT-S International Microwave Symposium}.\hskip 1em plus 0.5em minus 0.4em\relax IEEE, 2010, pp. 1266--1269.

\bibitem{milback}
\BIBentryALTinterwordspacing
H.~Lu, M.~Mazaheri, R.~Rezvani, and O.~Abari, ``A millimeter wave backscatter network for two-way communication and localization,'' in \emph{Proceedings of the ACM SIGCOMM 2023 Conference}, ser. ACM SIGCOMM '23.\hskip 1em plus 0.5em minus 0.4em\relax New York, NY, USA: Association for Computing Machinery, 2023, p. 49–61. [Online]. Available: \url{https://doi.org/10.1145/3603269.3604873}
\BIBentrySTDinterwordspacing

\bibitem{nesic1995frequency}
A.~Nesic and S.~Dragas, ``Frequency scanning printed array antenna,'' in \emph{IEEE Antennas and Propagation Society International Symposium. 1995 Digest}, vol.~2.\hskip 1em plus 0.5em minus 0.4em\relax IEEE, 1995, pp. 950--953.

\bibitem{xu2019wide}
S.-D. Xu, D.-F. Guan, Q.~Zhang, P.~You, S.~Ge, X.-X. Hou, Z.-B. Yang, and S.-W. Yong, ``A wide-angle narrowband leaky-wave antenna based on substrate integrated waveguide-spoof surface plasmon polariton structure,'' \emph{IEEE Antennas and Wireless Propagation Letters}, vol.~18, no.~7, pp. 1386--1389, 2019.

\bibitem{boskovic2017printed}
N.~Boskovic, B.~Jokanovic, and M.~Radovanovic, ``Printed frequency scanning antenna arrays with enhanced frequency sensitivity and sidelobe suppression,'' \emph{IEEE Transactions on Antennas and Propagation}, vol.~65, no.~4, pp. 1757--1764, 2017.

\bibitem{tutelian2021ieee}
S.~Tutelian, D.~Bankov, D.~Shmelkin, and E.~Khorov, ``Ieee 802.11 ax ofdma resource allocation with frequency-selective fading,'' \emph{Sensors}, vol.~21, no.~18, p. 6099, 2021.

\bibitem{WiFi-RTT}
``Wi-fi location: ranging with rtt,'' \url{https://developer.android.com/guide/topics/connectivity/wifi-rtt}, accessed: 2023-09-16.

\bibitem{IEEE802.11-2016}
``Ieee standard for information technology--telecommunications and information exchange between systems local and metropolitan area networks--specific requirements - part 11: Wireless lan medium access control (mac) and physical layer (phy) specifications,'' \url{https://standards.ieee.org/ieee/802.11/5536/}, accessed: 2023-09-16.

\bibitem{abedi2023wifi}
A.~Abedi, H.~Lu, A.~Chen, C.~Liu, and O.~Abari, ``Wifi physical layer stays awake and responds when it should not,'' \emph{IEEE Internet of Things Journal}, 2023.

\bibitem{CST}
{Dassault Systèmes}, ``{CST STUDIO SUITE},'' {Computer software}, 2023, {Available: https://www.3ds.com/products-services/simulia/products/cst-studio-suite/}.

\bibitem{zimaboard}
``Zimaboard,'' \url{https://www.zimaboard.com/}.

\bibitem{jiang2021eliminating}
Z.~Jiang, T.~H. Luan, X.~Ren, D.~Lv, H.~Hao, J.~Wang, K.~Zhao, W.~Xi, Y.~Xu, and R.~Li, ``Eliminating the barriers: Demystifying wi-fi baseband design and introducing the picoscenes wi-fi sensing platform,'' \emph{IEEE Internet of Things Journal}, vol.~9, no.~6, pp. 4476--4496, 2021.

\bibitem{MCSIndex2023}
\BIBentryALTinterwordspacing
{MCS Index}, ``Mcs index table, modulation and coding scheme index 11n, 11ac, and 11ax,'' 2023, accessed on yyyy-mm-dd. [Online]. Available: \url{https://mcsindex.com/}
\BIBentrySTDinterwordspacing

\bibitem{molisch2012wireless}
A.~F. Molisch, \emph{Wireless communications}.\hskip 1em plus 0.5em minus 0.4em\relax John Wiley \& Sons, 2012, vol.~34.

\bibitem{wirelessinsite}
{Remcom Inc.}, ``{Wireless InSite},'' {Computer software}, 2023, {Available: https://www.remcom.com/wireless-insite-em-propagation-software}.

\bibitem{wisight}
``Anonymized workshop paper.''

\bibitem{lwa}
D.~R. Jackson, C.~Caloz, and T.~Itoh, ``Leaky-wave antennas,'' \emph{Proceedings of the IEEE}, vol. 100, no.~7, pp. 2194--2206, 2012.

\bibitem{saeidi2021thz}
H.~Saeidi, S.~Venkatesh, X.~Lu, and K.~Sengupta, ``Thz prism: One-shot simultaneous localization of multiple wireless nodes with leaky-wave thz antennas and transceivers in cmos,'' \emph{IEEE Journal of Solid-State Circuits}, vol.~56, no.~12, pp. 3840--3854, 2021.

\bibitem{ghasempour2020single}
Y.~Ghasempour, C.-Y. Yeh, R.~Shrestha, D.~Mittleman, and E.~Knightly, ``Single shot single antenna path discovery in thz networks,'' in \emph{Proceedings of the 26th Annual International Conference on Mobile Computing and Networking}, 2020, pp. 1--13.

\bibitem{li20225g}
T.~Li, M.~H. Mazaheri, and O.~Abari, ``5g in the sky: the future of high-speed internet via unmanned aerial vehicles,'' in \emph{Proceedings of the 23rd Annual International Workshop on Mobile Computing Systems and Applications}, 2022, pp. 116--122.

\bibitem{fsawifi_gil2021wi}
A.~Gil-Mart{\'\i}nez, M.~Poveda-Garc{\'\i}a, J.~A. L{\'o}pez-Pastor, J.~C. S{\'a}nchez-Aarnoutse, and J.~L. G{\'o}mez-Tornero, ``Wi-fi direction finding with frequency-scanned antenna and channel-hopping scheme,'' \emph{IEEE Sensors Journal}, vol.~22, no.~6, pp. 5210--5222, 2021.

\bibitem{fsawifi_paaso2017doa}
H.~Paaso, N.~Gulati, D.~Patron, A.~Hakkarainen, J.~Werner, K.~R. Dandekar, M.~Valkama, and A.~M{\"a}mmel{\"a}, ``Doa estimation using compact crlh leaky-wave antennas: Novel algorithms and measured performance,'' \emph{IEEE Transactions on Antennas and Propagation}, vol.~65, no.~9, pp. 4836--4849, 2017.

\bibitem{fsawifi_poveda2020rssi}
M.~Poveda-Garc{\'\i}a, A.~G{\'o}mez-Alcaraz, D.~Ca{\~n}ete-Rebenaque, A.~S. Martinez-Sala, and J.~L. G{\'o}mez-Tornero, ``Rssi-based direction-of-departure estimation in bluetooth low energy using an array of frequency-steered leaky-wave antennas,'' \emph{IEEE Access}, vol.~8, pp. 9380--9394, 2020.

\bibitem{fsawifi_gil2023monopulse}
A.~Gil-Martinez, J.~A. L{\'o}pez-Pastor, M.~Poveda-Garc{\'\i}a, A.~Algaba-Braz{\'a}lez, D.~Ca{\~n}ete-Rebenaque, and J.~L. G{\'o}mez-Tornero, ``Monopulse leaky-wave antennas for rssi-based direction finding in wireless local area networks,'' \emph{IEEE Transactions on Antennas and Propagation}, 2023.

\bibitem{liu2022integrated}
F.~Liu, Y.~Cui, C.~Masouros, J.~Xu, T.~X. Han, Y.~C. Eldar, and S.~Buzzi, ``Integrated sensing and communications: Towards dual-functional wireless networks for 6g and beyond,'' \emph{IEEE journal on selected areas in communications}, 2022.

\bibitem{isac-survey}
K.~V. Mishra, M.~Bhavani~Shankar, V.~Koivunen, B.~Ottersten, and S.~A. Vorobyov, ``Toward millimeter-wave joint radar communications: A signal processing perspective,'' \emph{IEEE Signal Processing Magazine}, vol.~36, no.~5, pp. 100--114, 2019.

\bibitem{liu2018toward}
F.~Liu, L.~Zhou, C.~Masouros, A.~Li, W.~Luo, and A.~Petropulu, ``Toward dual-functional radar-communication systems: Optimal waveform design,'' \emph{IEEE Transactions on Signal Processing}, vol.~66, no.~16, pp. 4264--4279, 2018.

\bibitem{liu2020joint}
X.~Liu, T.~Huang, N.~Shlezinger, Y.~Liu, J.~Zhou, and Y.~C. Eldar, ``Joint transmit beamforming for multiuser mimo communications and mimo radar,'' \emph{IEEE Transactions on Signal Processing}, vol.~68, pp. 3929--3944, 2020.

\bibitem{sturm2011waveform}
C.~Sturm and W.~Wiesbeck, ``Waveform design and signal processing aspects for fusion of wireless communications and radar sensing,'' \emph{Proceedings of the IEEE}, vol.~99, no.~7, pp. 1236--1259, 2011.

\bibitem{he2023sencom}
Y.~He, J.~Liu, M.~Li, G.~Yu, J.~Han, and K.~Ren, ``Sencom: Integrated sensing and communication with practical wifi,'' in \emph{Proceedings of the 29th Annual International Conference on Mobile Computing and Networking}, 2023, pp. 1--16.

\end{thebibliography}





 





\end{document}